\documentclass[twocolumn]{aa}
  
\usepackage{graphicx}
\usepackage{amsmath,amsfonts,amssymb}
\usepackage{txfonts}
\usepackage{color}
\usepackage{natbib}
\usepackage{float}
\usepackage{dblfloatfix}
\usepackage{afterpage}
\usepackage{ifthen}
\usepackage[morefloats=12]{morefloats}
\usepackage{lipsum}
\usepackage{placeins}
\usepackage{subcaption}
\usepackage{capt-of}
\bibpunct{(}{)}{;}{a}{}{,}
\usepackage[switch]{lineno}
\definecolor{linkcolor}{rgb}{0.6,0,0}
\definecolor{citecolor}{rgb}{0,0,0.75}
\definecolor{urlcolor}{rgb}{0.12,0.46,0.7}
\usepackage[breaklinks, colorlinks, urlcolor=urlcolor,
    linkcolor=linkcolor,citecolor=citecolor,pdfencoding=auto]{hyperref}
\hypersetup{linktocpage}

\def\setsymbol#1#2{\expandafter\def\csname #1\endcsname{#2}}
\def\getsymbol#1{\csname #1\endcsname}

\def\Planck{\textit{Planck}}

\newbox\tablebox    \newdimen\tablewidth
\def\leaderfil{\leaders\hbox to 5pt{\hss.\hss}\hfil}
\def\endPlancktable{\tablewidth=\columnwidth 
    $$\hss\copy\tablebox\hss$$
    \vskip-\lastskip\vskip -2pt}

\def\tablenote#1 #2\par{\begingroup \parindent=0.8em
    \abovedisplayshortskip=0pt\belowdisplayshortskip=0pt
    \noindent
    $$\hss\vbox{\hsize\tablewidth \hangindent=\parindent \hangafter=1 \noindent
    \hbox to \parindent{$^#1$\hss}\strut#2\strut\par}\hss$$
    \endgroup}
\def\doubleline{\vskip 3pt\hrule \vskip 1.5pt \hrule \vskip 5pt}

\def\L2{\ifmmode L_2\else $L_2$\fi}

\def\DeltaT{\ifmmode \Delta T\else $\Delta T$\fi}
\def\deltat{\ifmmode \Delta t\else $\Delta t$\fi}
\def\fknee{\ifmmode f_{\rm knee}\else $f_{\rm knee}$\fi}
\def\Fmax{\ifmmode F_{\rm max}\else $F_{\rm max}$\fi}
\def\solar{\ifmmode{\rm M}_{\mathord\odot}\else${\rm M}_{\mathord\odot}$\fi}
\def\Msolar{\ifmmode{\rm M}_{\mathord\odot}\else${\rm M}_{\mathord\odot}$\fi}
\def\Lsolar{\ifmmode{\rm L}_{\mathord\odot}\else${\rm L}_{\mathord\odot}$\fi}
\def\inv{\ifmmode^{-1}\else$^{-1}$\fi}
\def\mo{\ifmmode^{-1}\else$^{-1}$\fi}
\def\sup#1{\ifmmode ^{\rm #1}\else $^{\rm #1}$\fi}
\def\expo#1{\ifmmode \times 10^{#1}\else $\times 10^{#1}$\fi}
\def\,{\thinspace}
\def\lsim{\mathrel{\raise .4ex\hbox{\rlap{$<$}\lower 1.2ex\hbox{$\sim$}}}}
\def\gsim{\mathrel{\raise .4ex\hbox{\rlap{$>$}\lower 1.2ex\hbox{$\sim$}}}}

\def\simprop{\mathrel{\raise .4ex\hbox{\rlap{$\propto$}\lower 1.2ex\hbox{$\sim$}}}}
\def\deg{\ifmmode^\circ\else$^\circ$\fi}
\def\pdeg{\ifmmode $\setbox0=\hbox{$^{\circ}$}\rlap{\hskip.11\wd0 .}$^{\circ}
          \else \setbox0=\hbox{$^{\circ}$}\rlap{\hskip.11\wd0 .}$^{\circ}$\fi}
\def\arcs{\ifmmode {^{\scriptstyle\prime\prime}}
          \else $^{\scriptstyle\prime\prime}$\fi}
\def\arcm{\ifmmode {^{\scriptstyle\prime}}
          \else $^{\scriptstyle\prime}$\fi}
\newdimen\sa  \newdimen\sb
\def\parcs{\sa=.07em \sb=.03em
     \ifmmode \hbox{\rlap{.}}^{\scriptstyle\prime\kern -\sb\prime}\hbox{\kern -\sa}
     \else \rlap{.}$^{\scriptstyle\prime\kern -\sb\prime}$\kern -\sa\fi}
\def\parcm{\sa=.08em \sb=.03em
     \ifmmode \hbox{\rlap{.}\kern\sa}^{\scriptstyle\prime}\hbox{\kern-\sb}
     \else \rlap{.}\kern\sa$^{\scriptstyle\prime}$\kern-\sb\fi}
\def\ra[#1 #2 #3.#4]{#1\sup{h}#2\sup{m}#3\sup{s}\llap.#4}
\def\dec[#1 #2 #3.#4]{#1\deg#2\arcm#3\arcs\llap.#4}
\def\deco[#1 #2 #3]{#1\deg#2\arcm#3\arcs}
\def\rra[#1 #2]{#1\sup{h}#2\sup{m}}

\def\dots{\relax\ifmmode \ldots\else $\ldots$\fi}
\def\WHzsr{\ifmmode $W\,Hz\mo\,sr\mo$\else W\,Hz\mo\,sr\mo\fi}
\def\mHz{\ifmmode $\,mHz$\else \,mHz\fi}
\def\GHz{\ifmmode $\,GHz$\else \,GHz\fi}
\def\mKs{\ifmmode $\,mK\,s$^{1/2}\else \,mK\,s$^{1/2}$\fi}
\def\muKs{\ifmmode \,\mu$K\,s$^{1/2}\else \,$\mu$K\,s$^{1/2}$\fi}
\def\muKRJs{\ifmmode \,\mu$K$_{\rm RJ}$\,s$^{1/2}\else \,$\mu$K$_{\rm RJ}$\,s$^{1/2}$\fi}
\def\muKHz{\ifmmode \,\mu$K\,Hz$^{-1/2}\else \,$\mu$K\,Hz$^{-1/2}$\fi}
\def\MJysr{\ifmmode \,$MJy\,sr\mo$\else \,MJy\,sr\mo\fi}
\def\MJysrmK{\ifmmode \,$MJy\,sr\mo$\,mK$_{\rm CMB}\mo\else \,MJy\,sr\mo\,mK$_{\rm CMB}\mo$\fi}
\def\microns{\ifmmode \,\mu$m$\else \,$\mu$m\fi}

\def\muK{\ifmmode \,\mu$K$\else \,$\mu$\hbox{K}\fi}
\def\microK{\ifmmode \,\mu$K$\else \,$\mu$\hbox{K}\fi}
\def\muW{\ifmmode \,\mu$W$\else \,$\mu$\hbox{W}\fi}
\def\kms{\ifmmode $\,km\,s$^{-1}\else \,km\,s$^{-1}$\fi}
\def\kmsMpc{\ifmmode $\,\kms\,Mpc\mo$\else \,\kms\,Mpc\mo\fi}
\providecommand{\sorthelp}[1]{}

\def\WMAP{\textit{WMAP}}

\def\commander{\texttt{Commander}}

\renewcommand{\d}[0]{\vec{d}}

\newcommand{\n}[0]{\vec{n}}

\definecolor{orange}{RGB}{255,127,0}

\newcommand{\s}[0]{\vec{s}}
\renewcommand{\a}[0]{\vec{a}}
\newcommand{\m}[0]{\vec{m}}

\newcommand{\B}[0]{\tens{B}}

\renewcommand{\L}[0]{\tens{L}}

\newcommand{\N}[0]{\tens{N}}

\newcommand{\M}[0]{\tens{M}}

\renewcommand{\r}[0]{\vec{r}}

\renewcommand{\P}[0]{\tens{P}}

\newcommand{\Dbp}[0]{\Delta_{\mathrm{bp}}}

\newcommand{\BP}{\textsc{BeyondPlanck}}

\def\inv{^{-1}}

\begin{document}

\title{\bfseries{\scshape{BeyondPlanck}} X. Bandpass and beam leakage corrections}
\newcommand{\nersc}[0]{1}
\newcommand{\princeton}[0]{2}
\newcommand{\helsinkiA}[0]{3}
\newcommand{\milanoA}[0]{4}
\newcommand{\triesteA}[0]{5}
\newcommand{\haverford}[0]{6}
\newcommand{\helsinkiB}[0]{7}
\newcommand{\triesteB}[0]{8}
\newcommand{\milanoB}[0]{9}
\newcommand{\milanoC}[0]{10}
\newcommand{\oslo}[0]{11}
\newcommand{\jpl}[0]{12}
\newcommand{\mpa}[0]{13}
\newcommand{\planetek}[0]{14}
\newcommand{\sandiego}[0]{15}
\author{\small
 T.~L.~Svalheim\inst{\oslo}\thanks{Corresponding author: T.~L.~Svalheim; \url{t.l.svalheim@astro.uio.no}},
K.~J.~Andersen\inst{\oslo}
\and
\textcolor{black}{R.~Aurlien}\inst{\oslo}
\and
\textcolor{black}{R.~Banerji}\inst{\oslo}
\and
M.~Bersanelli\inst{\milanoA, \milanoB, \milanoC}
\and
S.~Bertocco\inst{\triesteB}
\and
M.~Brilenkov\inst{\oslo}
\and
M.~Carbone\inst{\planetek}
\and
L.~P.~L.~Colombo\inst{\milanoA}
\and
H.~K.~Eriksen\inst{\oslo}
\and
\textcolor{black}{M.~K.~Foss}\inst{\oslo}
\and
C.~Franceschet\inst{\milanoC}
\and
\textcolor{black}{U.~Fuskeland}\inst{\oslo}
\and
S.~Galeotta\inst{\triesteB}
\and
M.~Galloway\inst{\oslo}
\and
S.~Gerakakis\inst{\planetek}
\and
E.~Gjerl{\o}w\inst{\oslo}
\and
\textcolor{black}{B.~Hensley}\inst{\princeton}
\and
\textcolor{black}{D.~Herman}\inst{\oslo}
\and
M.~Iacobellis\inst{\planetek}
\and
M.~Ieronymaki\inst{\planetek}
\and
\textcolor{black}{H.~T.~Ihle}\inst{\oslo}
\and
J.~B.~Jewell\inst{\oslo}
\and
\textcolor{black}{A.~Karakci}\inst{\oslo}
\and
E.~Keih\"{a}nen\inst{\helsinkiA, \helsinkiB}
\and
R.~Keskitalo\inst{\nersc}
\and
G.~Maggio\inst{\triesteB}
\and
D.~Maino\inst{\milanoA, \milanoB, \milanoC}
\and
M.~Maris\inst{\triesteB}
\and
S.~Paradiso\inst{\milanoA}
\and
B.~Partridge\inst{\haverford}
\and
M.~Reinecke\inst{\mpa}
\and
A.-S.~Suur-Uski\inst{\helsinkiA, \helsinkiB}
\and
D.~Tavagnacco\inst{\triesteB, \triesteA}
\and
H.~Thommesen\inst{\oslo}
\and
D.~J.~Watts\inst{\oslo}
\and
I.~K.~Wehus\inst{\oslo}
\and
A.~Zacchei\inst{\triesteB}
\and
A.~Zonca\inst{\sandiego}
}
\institute{\small
Computational Cosmology Center, Lawrence Berkeley National Laboratory, Berkeley, California, U.S.A.\goodbreak
\and
Department of Astrophysical Sciences, Princeton University, Princeton, NJ 08544,
U.S.A.\goodbreak
\and
Department of Physics, Gustaf H\"{a}llstr\"{o}min katu 2, University of Helsinki, Helsinki, Finland\goodbreak
\and
Dipartimento di Fisica, Universit\`{a} degli Studi di Milano, Via Celoria, 16, Milano, Italy\goodbreak
\and
Dipartimento di Fisica, Universit\`{a} degli Studi di Trieste, via A. Valerio 2, Trieste, Italy\goodbreak
\and
Haverford College Astronomy Department, 370 Lancaster Avenue,
Haverford, Pennsylvania, U.S.A.\goodbreak
\and
Helsinki Institute of Physics, Gustaf H\"{a}llstr\"{o}min katu 2, University of Helsinki, Helsinki, Finland\goodbreak
\and
INAF - Osservatorio Astronomico di Trieste, Via G.B. Tiepolo 11, Trieste, Italy\goodbreak
\and
INAF/IASF Milano, Via E. Bassini 15, Milano, Italy\goodbreak
\and
INFN, Sezione di Milano, Via Celoria 16, Milano, Italy\goodbreak
\and
Institute of Theoretical Astrophysics, University of Oslo, Blindern, Oslo, Norway\goodbreak
\and
Jet Propulsion Laboratory, California Institute of Technology, 4800 Oak Grove Drive, Pasadena, California, U.S.A.\goodbreak
\and
Max-Planck-Institut f\"{u}r Astrophysik, Karl-Schwarzschild-Str. 1, 85741 Garching, Germany\goodbreak
\and
Planetek Hellas, Leoforos Kifisias 44, Marousi 151 25, Greece\goodbreak
\and
San Diego Supercomputer Center, University of California, San Diego, U.S.A.\goodbreak
}

\authorrunning{Planck Collaboration}
\titlerunning{Diffuse component separation}

\abstract{We discuss the treatment of bandpass and beam leakage corrections in
  the Bayesian \BP\ CMB analysis pipeline as applied to the \Planck\ LFI
  measurements. As a preparatory step, we first apply three corrections to the
  nominal LFI bandpass profiles including removal of a known systematic effect
  in the ground measuring equipment at 61\,GHz; smoothing of standing wave
  ripples; and edge regularization. The main net impact of these modifications
  is an overall shift in the 70\,GHz bandpass of +0.6\,GHz; we argue that any
  analysis of LFI data products, either from \Planck\ or \BP, should use these
  new bandpasses. In addition, we fit a single free bandpass parameter for each
  radiometer of the form $\Delta_i = \Delta_0 + \delta_i$, where $\Delta_0$
  represents an absolute frequency shift per frequency band and $\delta_i$ is a
  relative shift per detector. The absolute correction is only fitted at 30\,GHz
  with a full $\chi^2$-based likelihood, resulting in a correction of
  $\Delta_{30}=0.24\pm0.03\,$GHz. The relative corrections are fitted using a
  spurious map approach, fundamentally similar to the method pioneered by the
  \WMAP\ team, but without introducing many additional degrees of freedom. All
  bandpass parameters are sampled using a standard Metropolis sampler within the
  main \BP\ Gibbs chain, and bandpass uncertainties are thus propagated to all
  other data products in the analysis. In total, we find that our bandpass model
  significantly reduces leakage effects. For beam leakage corrections, we adopt
  the official \textit{Planck} LFI beam estimates without additional degrees of
  freedom, and only marginalize over the underlying sky model. We note that this
  is the first time leakage from beam mismatch has been included for
  \textit{Planck} LFI maps.  }

\keywords{Cosmology: observations, polarization,
  cosmic microwave background --- Methods: data analysis, statistical}

\maketitle

\section{Introduction}
\label{sec:introduction} 

The cosmic microwave background (CMB), first discovered by
\citet{penzias:1965} is one of the most important sources of
information in cosmology. The most recent full-sky measurements of
this signal were made by the \Planck\ satellite \citep{planck2016-l01}
from an orbit around the second Sun-Earth Lagrange point between 2009
and 2013, using two complementary instruments to observe the sky in
nine frequency bands between 30 and 857\,GHz. These measurements have
put strong constraints on a wide range of both cosmological parameters
and physical phenomena, and form one of the cornerstones of
contemporary cosmology.

Although the official \Planck\ data processing ended in 2020
\citep{planck2016-l01,npipe}, several open questions regarding low-level
instrumental effects in \Planck\ remained unanswered at that time. Addressing
these for the Low Frequency Instrument (LFI; \citealp{planck2016-l02}) is a main
motivation for the \BP\ project \citep{bp01}. The \BP\ machinery is unique in
that it processes raw time ordered data (TOD) into final cosmological and
astrophysical results within one single integrated end-to-end Bayesian analysis
framework. Its computational engine is \commander\
\citep{eriksen:2004,eriksen2008,bp03}, which was originally developed for
\Planck\ component separation purposes
\citep{planck2013-p06,planck2014-a12,planck2016-l04}, and uses Gibbs sampling
\citep{geman:1984} to draw samples from a large global posterior distribution.
The \BP\ project has generalized this code to also account for low-level data
processing and mapmaking, and thereby integrated the full analysis pipeline into
a self-consistent Bayesian framework. For a full description of the \BP\
project, we refer the interested reader to \citet{bp01}.

A defining theme for the \BP\ approach is a detailed statistical
exploration of the interplay between instrumental effects and
astrophysical foregrounds.  Particularly relevant in this respect are
spectral responses. LFI comprised 22 radiometer chain assemblies
(RCAs) grouped in three bands with nominal frequencies of 30, 44 and
70\,GHz. Within a band, each RCA has a slightly different spectral
response (or bandpass profile) and center frequency compared to the
others. The \Planck\ LFI spectral responses used in the official
analysis were provided as part of the 2018 data
release.\footnote{\url{https://pla.esac.esa.int/}} These response
functions were measured on the ground prior to launch, as described by
\citet{zonca2009}.  However, laboratory measurements of bandpass
profiles are in general a highly non-trivial task, as even small
environmental variations and interferences may affect the
results. Several systematic issues were identified during the LFI
testing campaign, and some of these were left uncorrected in the final
LFI products, even though approaches for possible improvements were
discussed. In this paper, we finally implement these corrections, and
make the resulting bandpass profiles publicly
available.\footnote{\url{https://beyondplanck.science/products/files/}}
We also recommend that any future analysis of the \Planck\ LFI data
should use the new bandpasses, as the effects are non-negligible and
lead to improved internal consistency.

Even with perfect knowledge of the instrument, there will be artefacts
in the final frequency and component maps when using a multi-detector
mapmaking algorithm (which includes those employed by \Planck;
\citealp{planck2016-l02,planck2016-l03}) if the detectors have
different sensitivities, unless these differences are properly
accounted for \citep[e.g.,][]{page2007,npipe,delouis:2019}. This
applies both to differences in bandpass and beam profiles. In this
paper, we discuss how such corrections are applied within the
\BP\ framework, both in terms of how to correct deterministically for
an assumed known response functions, and how to account for
uncertainties in the response functions themselves.

We note that for the white noise levels typical of \WMAP\ and \Planck, 
previous static leakage corrections suppress residual effects well below the
levels relevant for cosmological interpretation. However, future CMB experiments
will target primordial gravitational waves (e.g., \citep{kamionkowski:2016} and
references therein) and more specifically the tensor-to-scalar ratio, $r$.
Current measurements from BICEP2/Keck and \Planck\ constrain this parameter to
$r<0.036$ \citep{bicep2021} and $r<0.032$ \citep{tristram:2021}, respectively,
and these low levels correspond to signals that are only a few hundreds
of nanokelvin on the sky. To reach levels of $r\lesssim10^{-3}-10^{-4}$, highly
accurate modelling of bandpass and beam leakage effects will be critically
important, and we believe that an integrated method of the kind shown in this paper will be required.

The rest of this paper is organized as follows. We start by discussing
the LFI bandpass pre-processing in Sect.~\ref{sec:newprofs}. We then
present our own algorithms in Sect.~\ref{sec:methods}, and show how
these directly build on and generalize previous efforts. Results from
the main Markov chain analysis are reported in
Sect.~\ref{sec:results}, and we finally conclude in
Sect.~\ref{sec:conclusion}.

\section{Bandpass pre-processing}
\label{sec:newprofs}

\begin{figure*}[p]
  \center
  \begin{subfigure}{0.45\linewidth}
    \includegraphics[width=\linewidth]{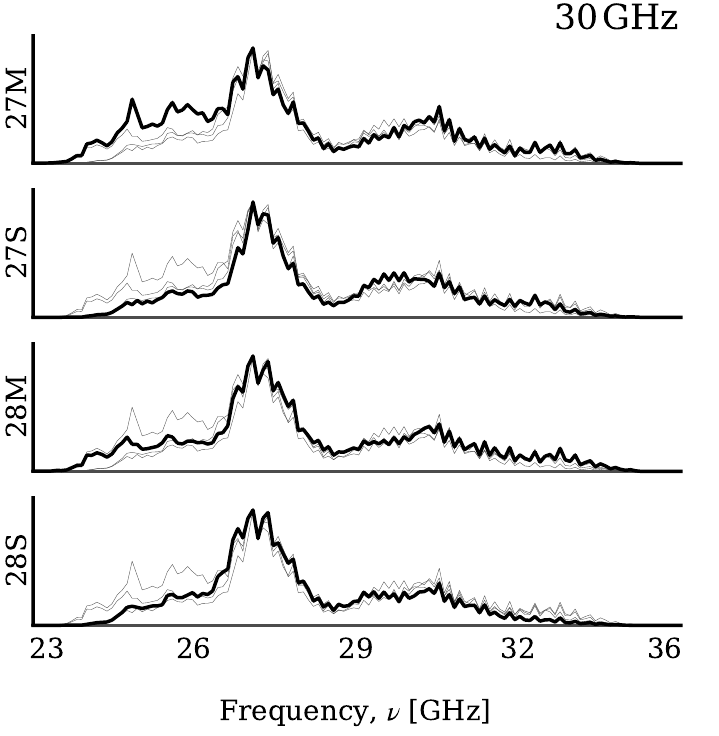}\vspace*{2.1cm}
    \includegraphics[width=\linewidth]{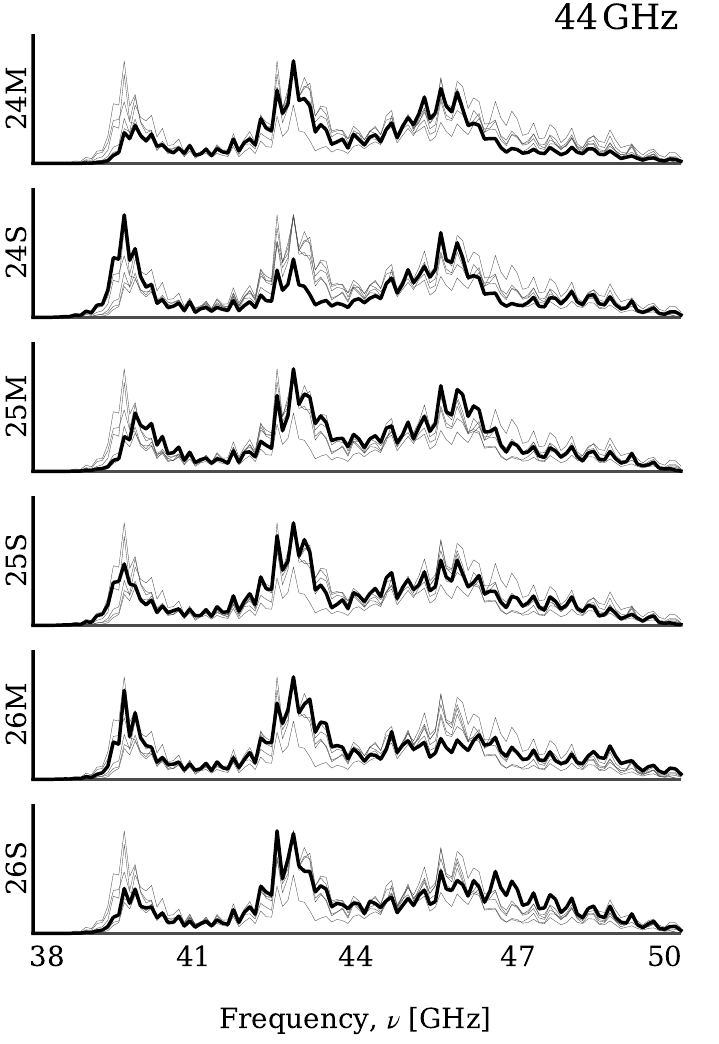}
  \end{subfigure}
  \begin{subfigure}{0.45\linewidth}
    \includegraphics[width=\linewidth]{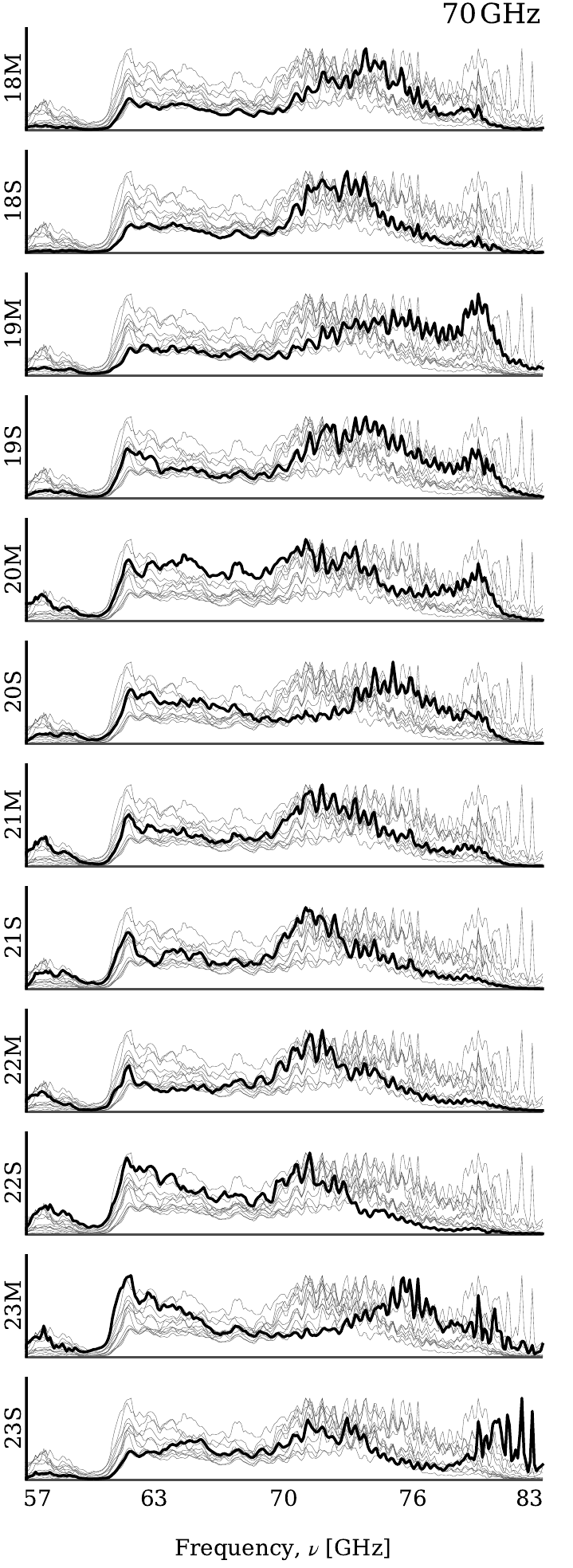}
  \end{subfigure}
  \caption{Normalized radiometer bandpass profiles for LFI 30\,GHz
    (\emph{top left}), 44\,GHz (\emph{bottom left}), and 70\,GHz
    (\emph{right}). Individual bandpasses are highlighted in bold and the remaining are plotted behind for ease of comparison.}
  \label{fig:bp_LFI}
\end{figure*}

We start our discussion by reviewing the bandpass profiles provided by the
official \Planck\ LFI data processing center (DPC) \citep{planck2013-p02}, as
shown in Fig.~\ref{fig:bp_LFI} for each radiometer. As summarized by
\citet{zonca2009}, these functions were characterized through two complementary
approaches. The first was to model and characterize each optical element
individually, and then combine these into a complete model. The second approach
was through integrated cryogenic system tests, which for LFI were carried out by
Thales Alenia Space in Vimodrone for the 30 and 44\,GHz channels, and by
DA-Design Ylinen in Finland for the 70\,GHz channel.

While the results from the two approaches generally agreed, there were also
notable differences, some of which may be identified by eye in
Fig.~\ref{fig:bp_LFI}. Perhaps the most striking feature is standing waves,
which result in high-frequency ripples across the bandpass. While some of these
may be due to real features in RCA itself, others may be caused by standing
waves excited by the testing equipment. One particularly notable example of this
is the 70\,GHz channel, for which the input load was placed directly in front of
the feed horn during the test, resulting in strong standing waves between the two components. At the same
time, the precise phase of the ripples is sensitive to environmental properties,
and can for instance change depending on the ambient temperature. As such, it is
non-trivial to assign a physical reality to these ripples as far as the real
measurements are concerned. \citet{zonca2009} therefore suggested that these
should be removed through low-pass filtering before cosmological data analysis.
This step was, however, never actually implemented during the official \Planck\
analysis, and we therefore do this in this paper. Technically speaking, we
implement this filter in logarithmic space using a 3-pole low pass filter at 0.2
times the Nyquist frequency. We note, however, that the specific details
regarding the filter are not critically important.

\begin{figure*}[p]
  \center
  \begin{subfigure}{0.45\linewidth}
    \includegraphics[width=\linewidth]{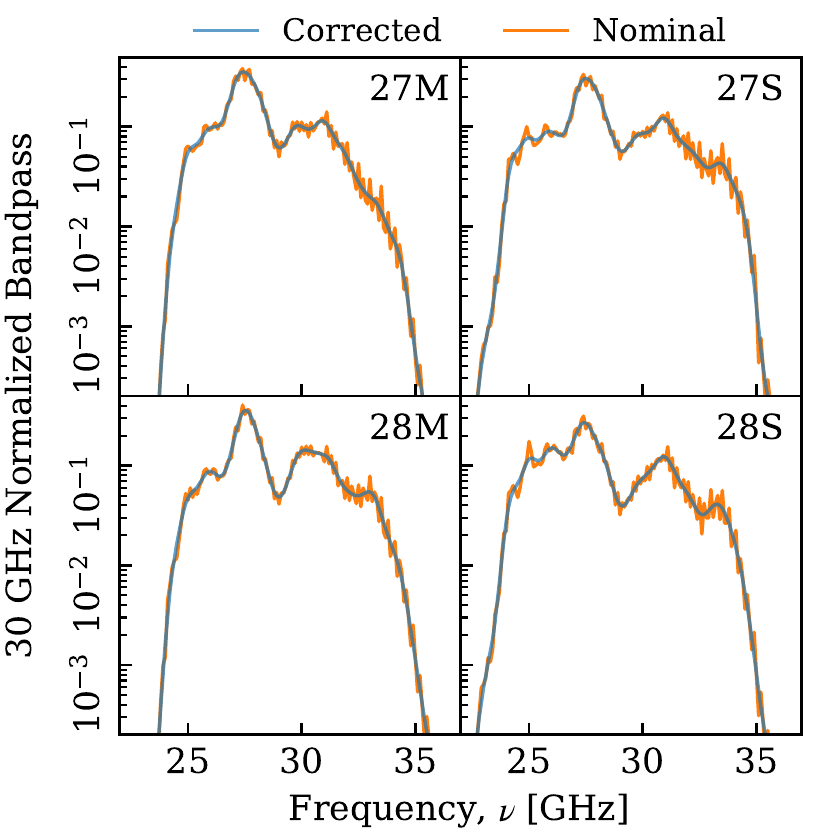}\vspace*{2.cm}
    \includegraphics[width=\linewidth]{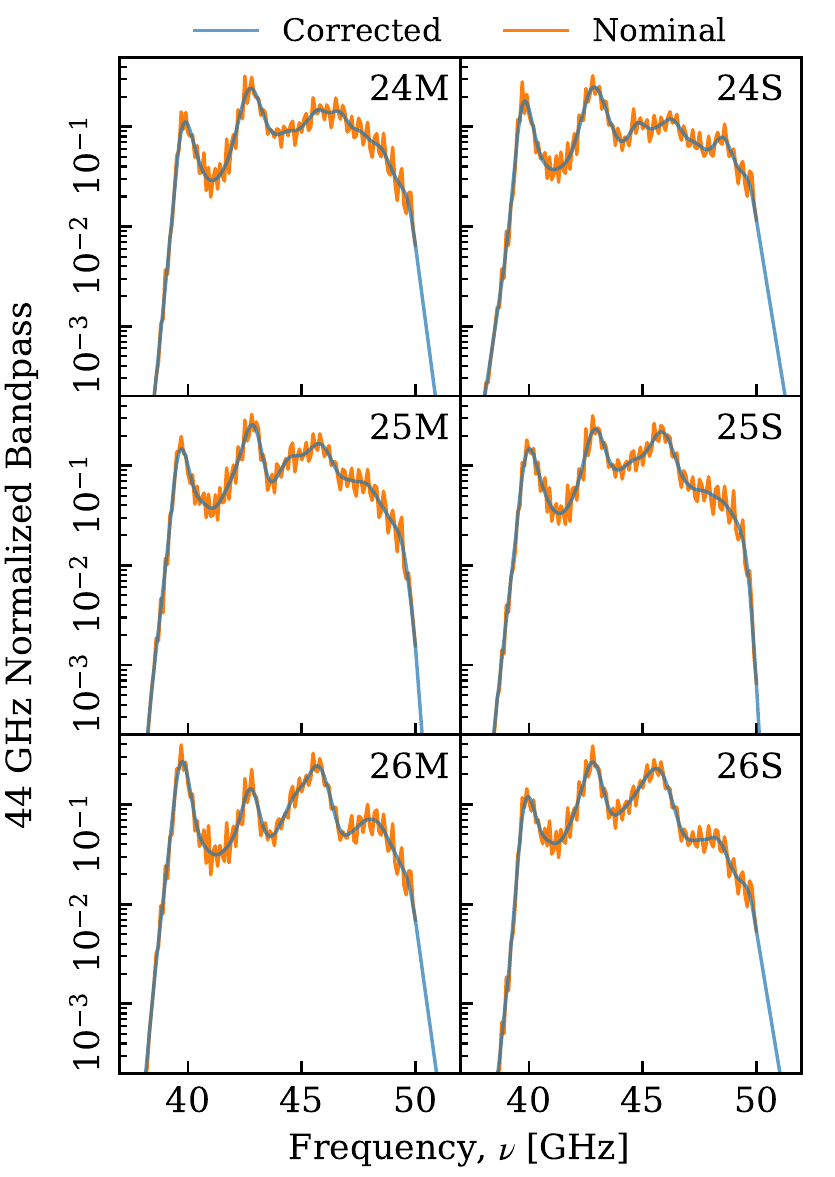}
  \end{subfigure}
  \begin{subfigure}{0.45\linewidth}
    \includegraphics[width=\linewidth]{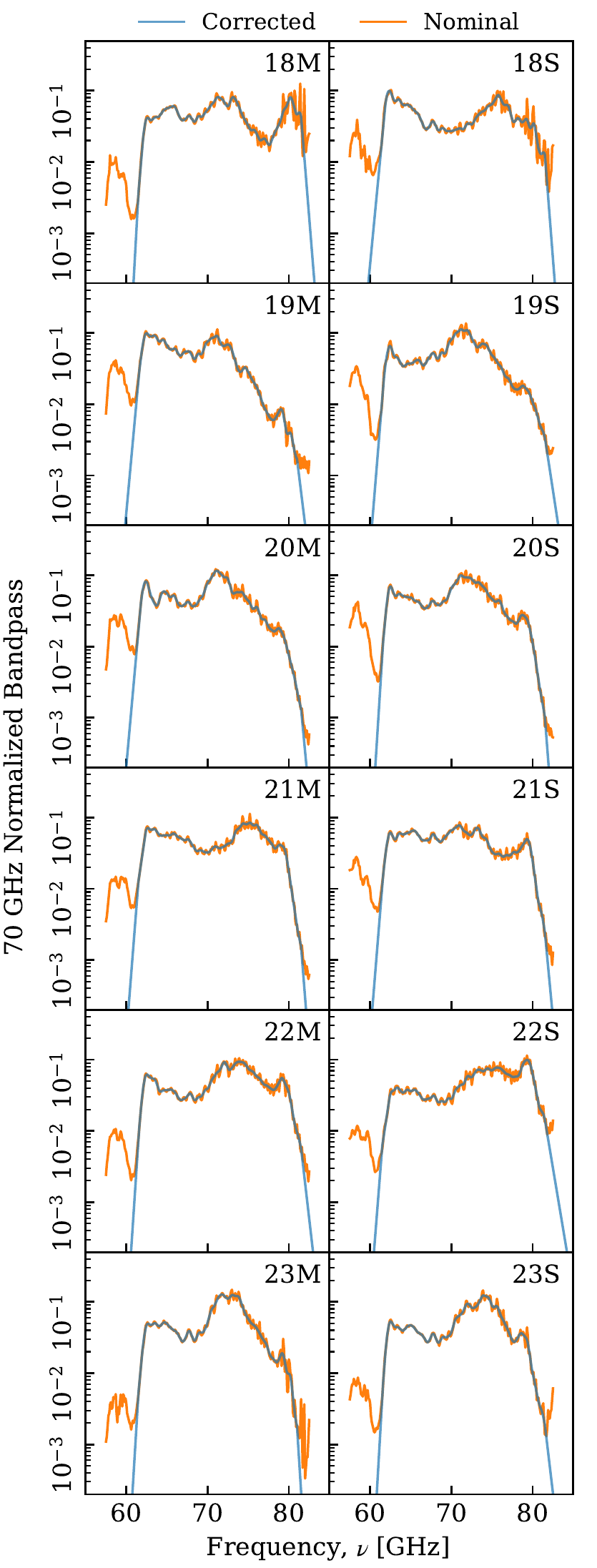}
  \end{subfigure}
  \caption{Nominal and corrected bandpass profiles for LFI 30\,GHz
    (\emph{top left}), 44\,GHz (\emph{bottom left}), and 70\,GHz
    (\emph{right}) radiometers. Orange curves show raw bandpass
    profiles as provided by the \Planck\ LFI DPC, and blue curves show
  the corrected profiles discussed in Sect.~\ref{sec:newprofs}. }
  \label{fig:corrected_bp}
\end{figure*}

A second known artefact induced by the testing equipment is the excess
seen in the 70\,GHz bandpasses below 61\,GHz, which is due to a known
systematic effect in all gain measurements of the backend module
(BEM); it is thus caused by the test equipment, and not the
radiometers themselves \citep{zonca2009}. This feature should simply
be removed, and the 70\,GHz bandpass should be limited to 61--80\,GHz.

\begin{table}[t]
  \newdimen\tblskip \tblskip=5pt
  \caption{Effective center frequencies, $\nu_{\rm eff}$, and unit conversion factors, $U_c$, for nominal and corrected bandpass profiles.\label{tab:bpcorr}}
  \vskip -4mm
  \footnotesize
  \setbox\tablebox=\vbox{
   \newdimen\digitwidth
   \setbox0=\hbox{\rm 0}
   \digitwidth=\wd0
   \catcode`*=\active
   \def*{\kern\digitwidth}
    \newdimen\dpwidth
    \setbox0=\hbox{.}
    \dpwidth=\wd0
    \catcode`!=\active
    \def!{\kern\dpwidth}
    \halign{\hbox to 1.2cm{$#$\leaderfil}\tabskip .5em&
      $#$\hfil \tabskip 0.5em&
      \hfil$#$\hfil \tabskip 0.5em&
    \hfil$#$\hfil \tabskip 0.5em&      
    \hfil$#$\hfil \tabskip 0.5em&
    \hfil$#$\hfil \tabskip 0em\cr
  \noalign{\doubleline}
  \omit\hfil\rm Quantity \hfil & \rm Unit & \rm Band & \rm Nominal &\rm Corrected &\rm Change![$\%$]\cr 
  \noalign{\vskip 3pt\hrule\vskip 5pt}
\nu_{\rm eff} &  \rm GHz & 30             &28.756&28.747&-0.03\cr 
\omit & \omit & 44               &44.121&44.148&*0.06\cr
\omit & \omit & 70               &70.354&70.795&*0.63\cr
  \noalign{\vskip 3pt} 
U_c & \rm K_{CMB}/K_{RJ} & 30            &1.0217&*1.0217&*0.00\cr
\omit & \omit & 44            &1.0515&*1.0515&*0.00\cr
\omit & \omit & 70            &1.1353&*1.1369&*0.14\cr
  \noalign{\vskip 3pt} 
U_c & \rm   MJy\, sr^{-1} /K_{CMB} & 30 &*24.269&*24.254&-0.06\cr
\omit & \omit & 44 &*55.723&*55.834&*0.19\cr
\omit & \omit & 70 &129.22*&132.45*&*2.49\cr
\noalign{\vskip 5pt\hrule\vskip 5pt}}}
\endPlancktable
\end{table}

More generally, the edges of the bandpasses are not well
characterized, and standing wave effects also tend to be relatively
larger near the edges. The bandpass cut-offs should therefore be
regularized through smooth apodization. In this paper, we implement
this by calculating the derivative of the low-pass filtered profiles
near the edges, and extrapolate smoothly to zero. For the 70\,GHz
channel, the derivative is evaluated at 61\,GHz to remove the BEM
artefact discussed above. 

Figure~\ref{fig:corrected_bp} shows a comparison of the raw and corrected
bandpasses for each radiometer. Here we see that the main change applied to the
30\,GHz radiometers is the low-pass filter, as the edges were already quite well
characterized in the original measurements. For the 44\,GHz band, the most
prominent correction is the high frequency cutoff. In particular, extrapolating
the behavior for horn 24 is not trivial, as the profiles do not show a
convincing downward trend before truncation. However, the profiles for both horn 25 and 26 suggest
that the power drop-off is most likely just outside the measured range. 

Edge trimming is also the dominant effect for the 70\,GHz channels,
both at low and high frequencies. The substantial amplitude and width
of the 61\,GHz spike feature implies that even the center frequency of
the 70\,GHz channel will change notably for this channel. This is
quantified in Table~\ref{tab:bpcorr}; see next section for exact definitions of each quantity. Here
we see that the 70\,GHz center frequency increases by 0.44\,GHz or,
equivalently, by 0.6\,\%, which translates into a difference in the
conversion factor between flux density and thermodynamic temperature
units, $U_c$, of 2.5\,\%. The impact of these corrections in terms of
data quality and astrophysical products is considered in
Sect.~\ref{sec:results}. The Python script that applies the various
corrections is available
online.\footnote{\url{https://github.com/trygvels/BeyondPlanck_LFI_bandpass}}

\begin{figure*}[t]
    \includegraphics[width=0.33\linewidth]{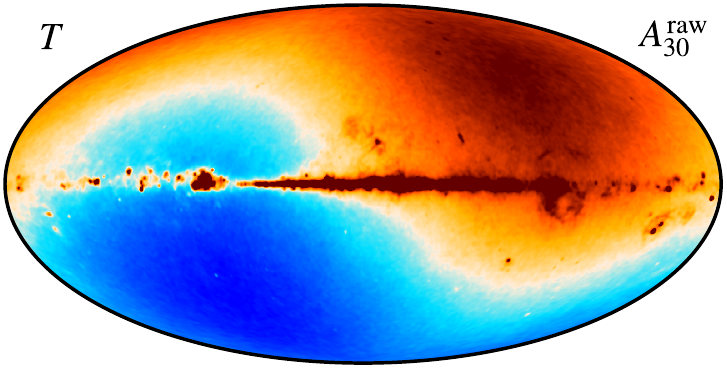}
    \includegraphics[width=0.33\linewidth]{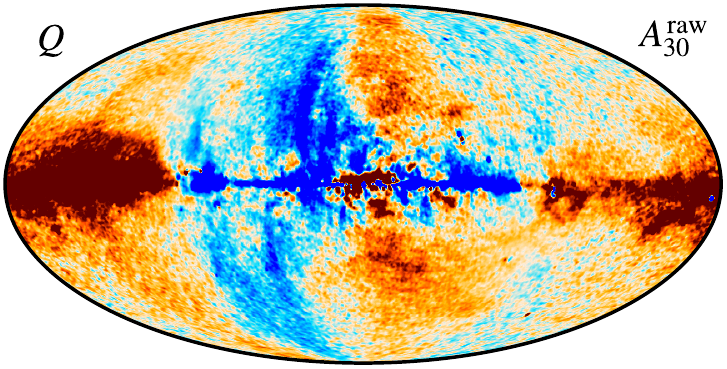}
    \includegraphics[width=0.33\linewidth]{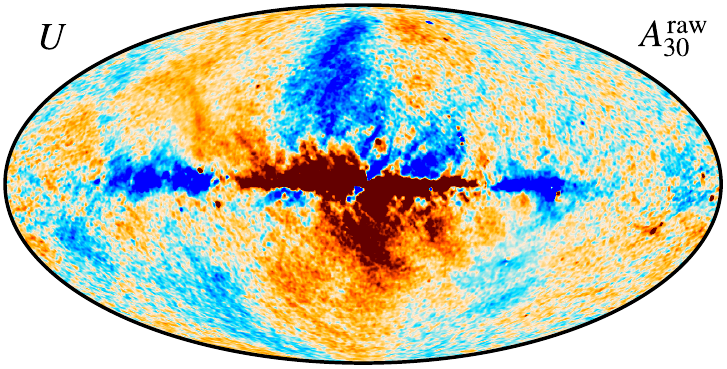}\\
    \includegraphics[width=0.33\linewidth]{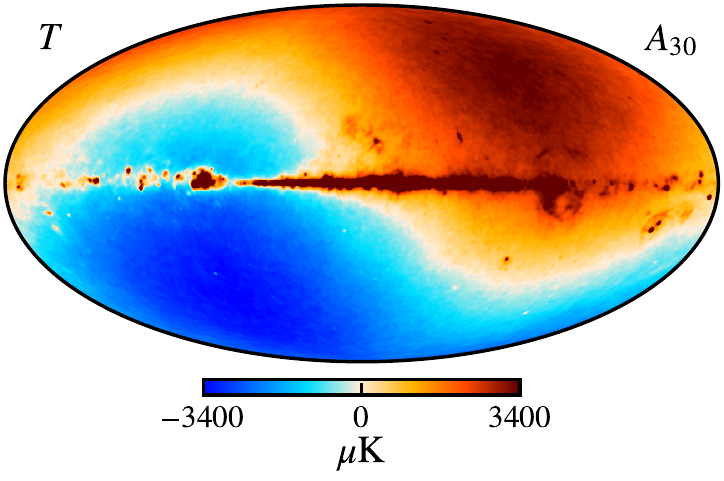}
    \includegraphics[width=0.33\linewidth]{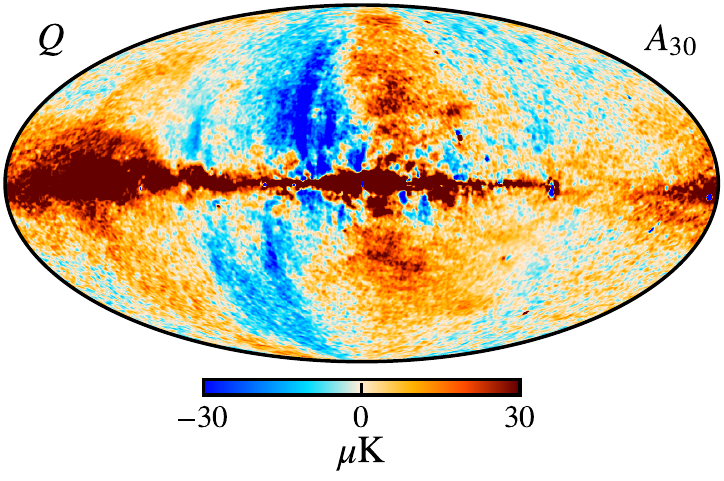}
    \includegraphics[width=0.33\linewidth]{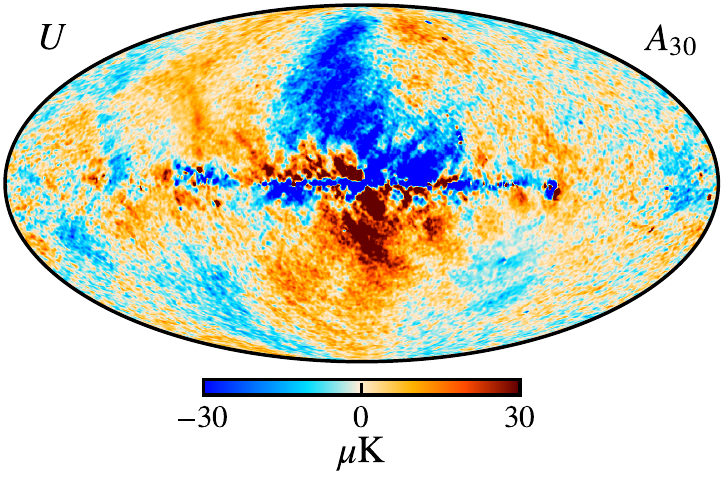}\\
    \includegraphics[width=0.33\linewidth]{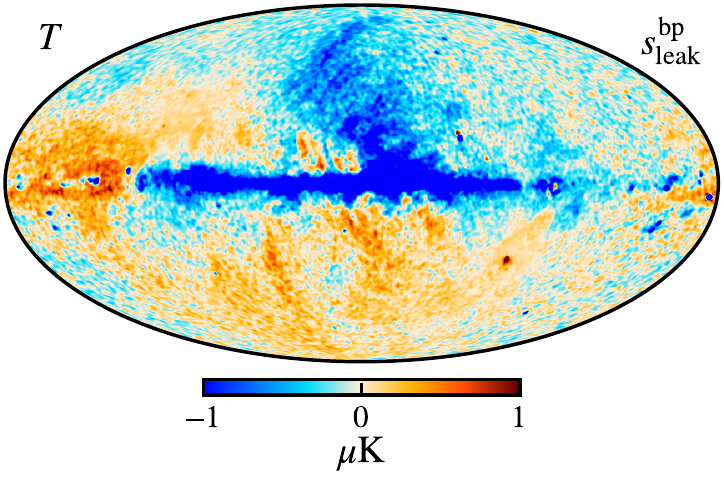}
    \includegraphics[width=0.33\linewidth]{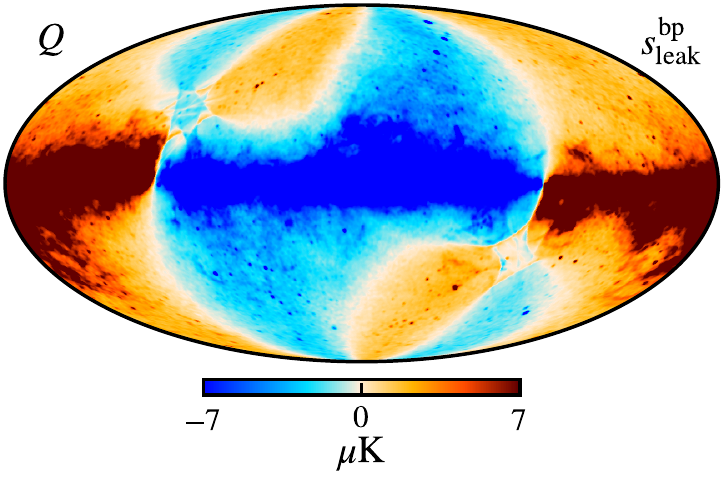}
    \includegraphics[width=0.33\linewidth]{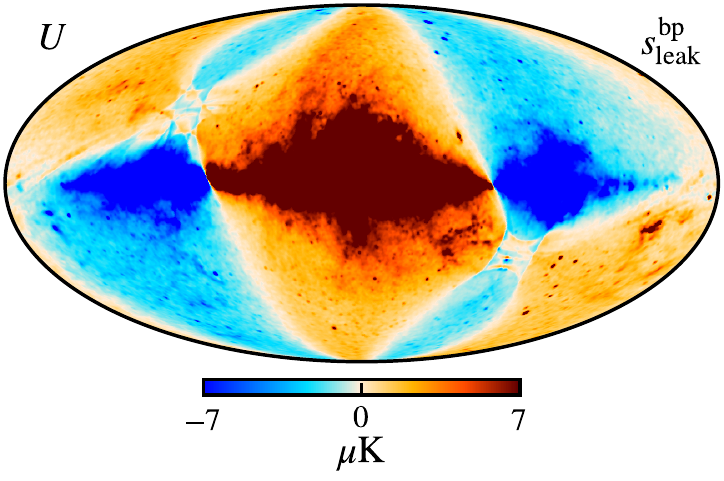}\\    
    \includegraphics[width=0.33\linewidth]{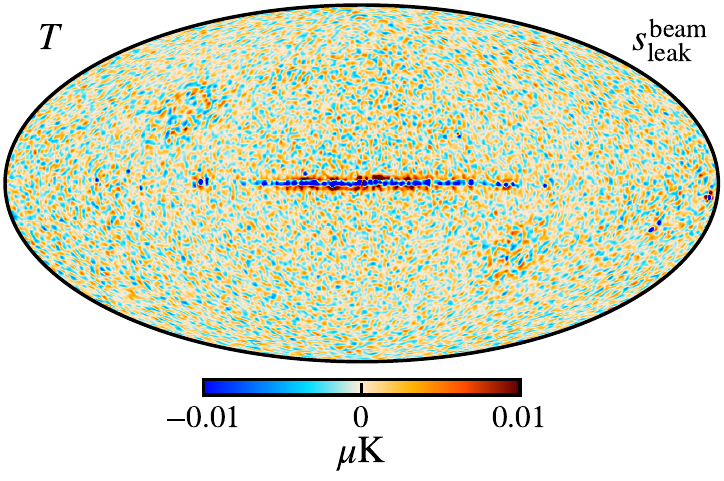}
    \includegraphics[width=0.33\linewidth]{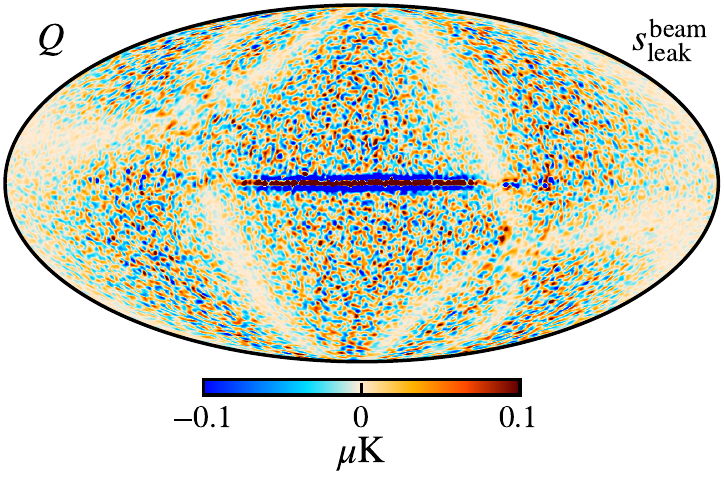}
    \includegraphics[width=0.33\linewidth]{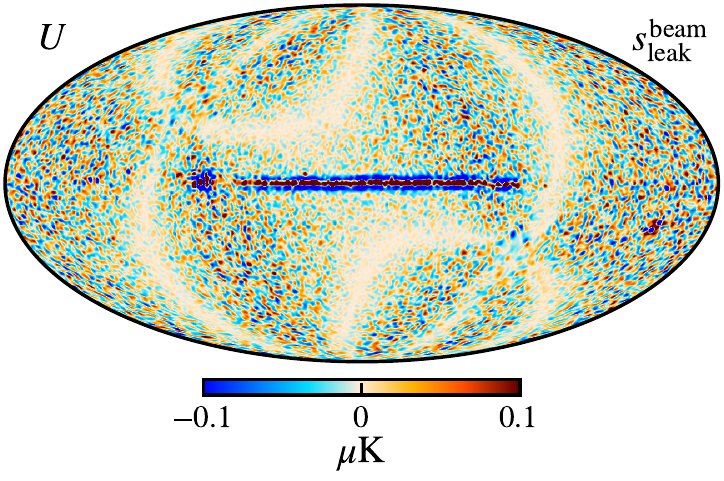}\\
  \caption{Comparison of the raw (\emph{top row}) and leakage corrected 
    (\emph{second row}) \Planck\ 30\,GHz channel. The third and fourth rows show
    the individual bandpass and beam leakage corrections,
    respectively. Columns shows Stokes $T$, $Q$, and $U$ parameters. All maps have been smoothed to an angular resolution of 1\deg\ FWHM, except for the bottom row, which has been smoothed to 2\deg\ FWHM. }\label{fig:subleakmaps_30}
\end{figure*}

\begin{figure}[t] 
  \includegraphics[trim=0 60 0 60, width=.49\linewidth]{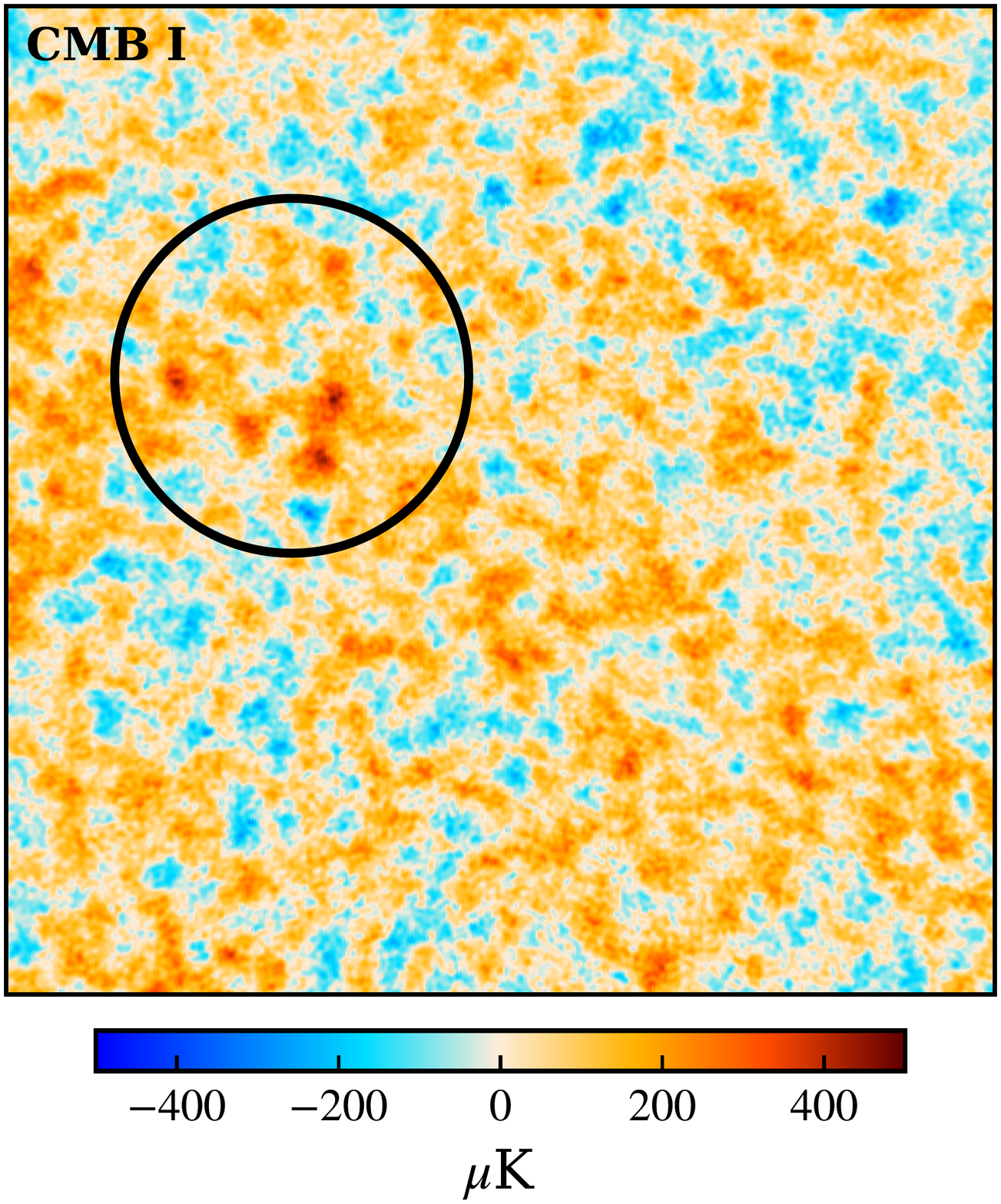}
  \includegraphics[trim=0 60 0 60, width=.49\linewidth]{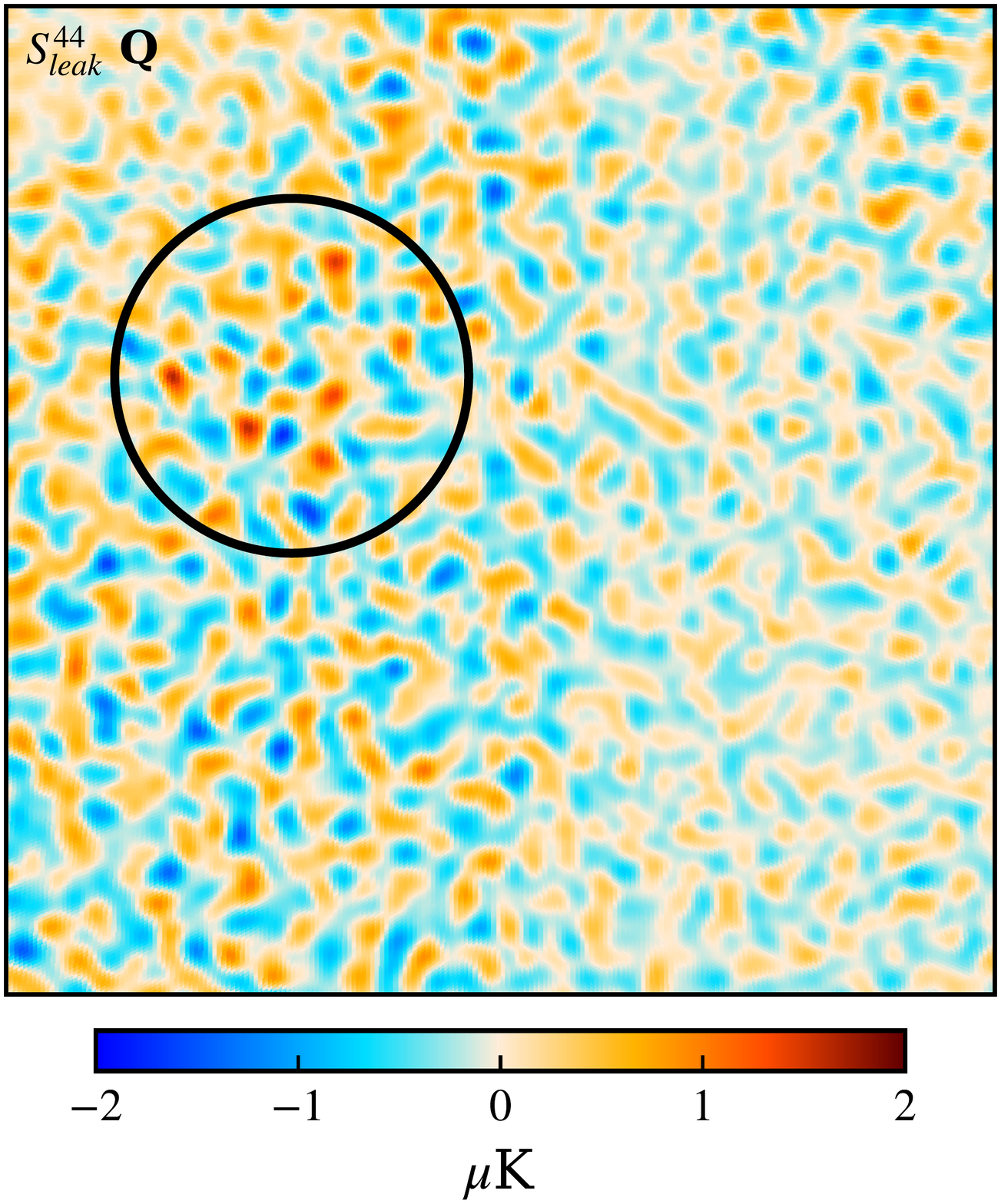}\\
  \includegraphics[width=\linewidth]{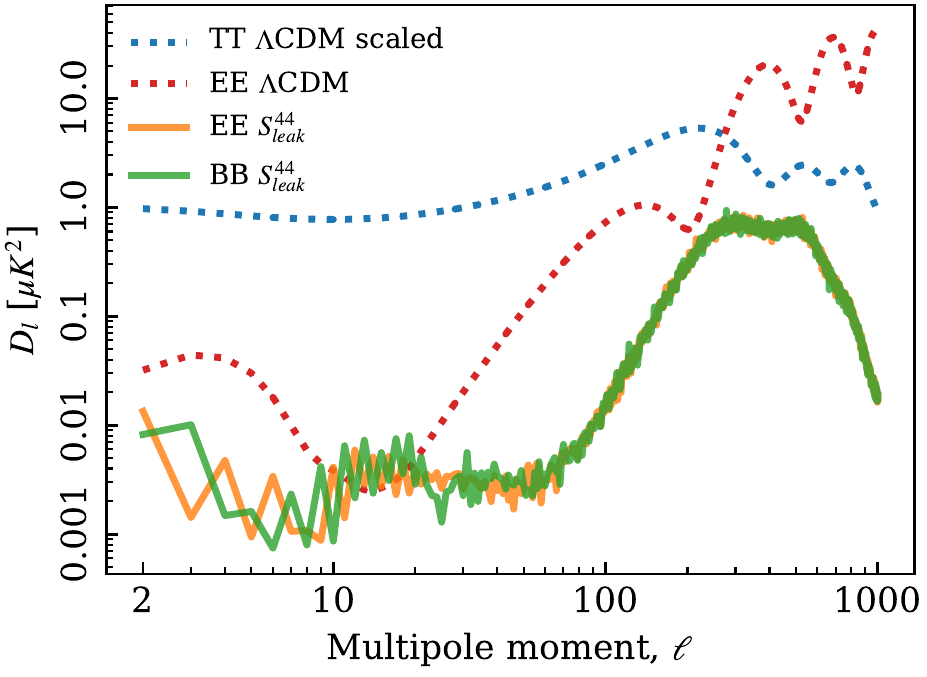}
  \caption{(\emph{Top left:}) Gnomonic projection of a
    $30^\circ\times30^\circ$ area centered on $(l,b)=(0\deg,-70\deg)$
    in the \BP\ posterior mean CMB temperature map
    \citep{bp11}. (\emph{Top right:}) \BP\ 44\,GHz $Q$-leakage map in
    the same area. (\emph{Bottom:}) Angular polarization power spectra
    of the masked 44\,GHz leakage map, compared with the
    \Planck\ best-fit $\Lambda$CDM $TT$ (scaled by $10^{-3}$) and $EE$
    spectra.}\label{fig:beamleakage}
\end{figure}

\section{Methodology}
\label{sec:methods}

The main goal of the \BP\ project is to perform an end-to-end Bayesian analysis
of the \Planck\ LFI data. To do so, we start by writing down a physically
motivated parametric model which can be fitted to our calibrated time-ordered
data, $\d$ for radiometer $j$,
\begin{equation}
  \d_j = \P_j\B_j\int \s(\nu)\tau_j(\nu)\,\mathrm d\nu + \n_j.
  \label{eq:todmodel}
\end{equation}
Here $\P$ is a pointing matrix that maps the sky signal $\s$ into time
domain; $\B$ is the beam profile; and $\tau(\nu)$ is the bandpass. The
sky signal as observed by radiometer $j$ may be written in the
following form,
\begin{equation}
  \s^j = \sum_{c=1}^{N_{\mathrm{comp}}}\a_c\, \left[\,U_j \int f_c(\nu; \beta)\,
    \tau_j(\nu)\,\mathrm d\nu\right] \equiv
  \sum_{c=1}^{N_{\mathrm{comp}}}\M^j_c\,\a_c = \M^j\,\a,
  \label{eq:mixmat}
\end{equation}
where the sum runs over all relevant astrophysical sky components,
$c$, each with its own unit conversion factor $U_j$ that converts from its own intrinsic unit to the common brightness temperature unit adopted for $\s$. As shown by \citep{planck2013-p03d}, this may be written as
\begin{equation}
U_{ij} = \frac{\int \tau(\nu)\frac{dI_{\nu}}{dX_i}\, \mathrm d\nu}{\int
  \tau(\nu)\frac{dI_{\nu}}{dX_j}\,\mathrm d\nu},
\end{equation}
where $dI_{\nu}/dX_i$ is the intensity derivative expressed in the given unit convention.

As defined by Eq.~\eqref{eq:mixmat}, $\M_c^j$ is called the \emph{mixing
  matrix}, and this matrix scales the set of component amplitudes $\a$ to
arbitrary frequencies using the spectral energy distribution $f$ of each component,
which may be described by some set of spectral parameters
$\beta$. Using this notation, Eq.~\eqref{eq:todmodel} be written in
the following slightly more compact form,
\begin{equation}
  \d_j = \P_j\B_j\M_j\a + \n_j.
  \label{eq:todmodelmat}
\end{equation}
For a full discussion of the \BP\ data model and notation, we
refer the interested reader to \citet{bp01} and references therein.

In this paper, we are particularly interested in how beam and bandpass
differences between detectors create leakage effects, and how to correct for
these. The first complication in that respect is related to binning the
time-ordered data, $\d$, into Stokes parameter maps, $\m_\nu$, using the mapmaking
equation,
\begin{equation}
  \left(\sum_{j \in \nu} \P_j^t \N_{j}^{-1} \P_j\right) \m_{\nu} =
  \sum_j \P_j^t \N_j^{-1}\d_j,
  \label{eq:binmap}
  \end{equation}
where we now have defined $\N_j=\left<\n\n^t\right>$ to be the
(time-domain) noise covariance matrix for detector $j$. 

This equation implicitly assumes that all radiometers, $j$, observe the same
sky, $\s$. However, although the various detectors actually do observe the same
sky, their \emph{response} to the spectral distribution of the different
foreground components varies, because the bandpass and beam profiles differ, as
described by Eq.~\eqref{eq:todmodel}. When co-adding these different
measurements into a sky map, any difference from the mean will be interpreted as
noise by the mapmaker, and accordingly be distributed between the various Stokes
parameters according to the local scanning and noise levels at any given time.
Furthermore, since these discrepancies are not stochastic, but deterministically
predictable by the bandpass and beam profiles, they do not average down in time.
They therefore induce systematic errors in the final maps that are directly
correlated with the sky signal itself. These effects are particularly
significant for polarization analysis, where different bands must be combined to
extract the very low Q and U signals. These errors are therefore particularly
worrisome for cosmological and astrophysical analyses.

We define three different effects caused by bandpass and beam
errors. The first effect is called \textit{bandpass mismatch}, and
this describes the deterministic effect discussed above, namely that
different bandpass profiles create spurious leakage during
multi-detector mapmaking, creating what is often referred to as
``temperature-to-polarization leakage''. However, this name is
somewhat of misnomer, since all Stokes parameters are formally coupled
\citep[see, e.g.,][]{npipe}.  Still, the effect is relatively much
more important for polarization than for temperature because of its
far lower signal-to-noise ratio. Indeed, this effect is the single
strongest instrumental polarization contaminant for the LFI 30\,GHz
channel (see, e.g., Fig.~16 in \citealp{bp01}), but, fortunately, it
is also entirely predictable, and could in principle be removed to machine
accuracy if both the astrophysical sky and the detector bandpasses
were perfectly known.

However, as discussed above, the bandpasses are by no means perfectly
known, and those uncertainties create additional leakage that is not
deterministically correctable.  Furthermore, since all radiometer
bandpasses within a band are uncertain, the combined co-added
frequency bandpass is also uncertain. And this uncertainty produces an
artefact when incorrectly translating the foreground sky model to the
observed signal through the mixing matrix in Eq.~\eqref{eq:mixmat}. We
refer to the effect caused by bandpass uncertainties as
\textit{bandpass errors}, and we attempt to minimize and marginalize
over these by parameterizing the bandpasses, and fit the associated
free parameters as part of the main Gibbs sampling process.

The third and final leakage effect considered here is \textit{beam mismatch},
which is a deterministic leakage effect arising from differences between the
main beams of the radiometers in a given frequency channel. These are in
principle similar to the bandpass mismatch effect, but typically only affect
small angular scales. In this paper we will only account for static FWHM
differences between radiometers, but not asymmetric beams---that will be
discussed in future publications, as will \emph{beam errors}, i.e.,
uncertainties in the actual beam profiles.

\subsection{Leakage corrections}
\label{sec:leakcorr}

As discussed above, bandpass mismatch artefacts arise because
Eq.~\eqref{eq:binmap} assumes that all radiometers measure the same signal in
any given pixel of the sky, while, they do not, because their different bandpass
and beam profiles couple differently with the SED of foreground emissions. To
account for these differences, we define the following leakage correction term
for each radiometer,
\begin{equation}
  \delta s^{\mathrm{leak}}_{j, t} = \P^{j}_{tp}\B_{pp'}^j\left(s^{\mathrm{sky}}_{jp'} - \left<s^{\mathrm{sky}}_{jp'}\right>\right).
 \label{eq:leak}
\end{equation}
Here, $s^{\mathrm{sky}}_{j}$ is a model of the sky as actually seen by
radiometer $j$ at pixel $p$, taking into account its specific bandpass and beam
profile, and angled brackets denote an average over all radiometers
evaluated pixel-by-pixel. Note that this leakage term explicitly
accounts for both bandpass and beam differences through
$s^{\mathrm{sky}}_{j}$ and $\B_{pp'}^j$.

In order to create a leakage-cleaned frequency map, we simply subtract
this leakage term from the calibrated data prior to mapmaking,
\begin{equation}
  \left(\sum_{j \in \nu} \P_j^t \N_{j}^{-1} \P_j\right) \m_{\nu} =
  \sum_{j \in \nu} \P_j^t \N_j^{-1}(\d_j -
  \delta\s^{\mathrm{leak}}_j).
  \label{eq:mapmaking_corr}
\end{equation}
In this equation, the right-hand side corresponds to a stationary sky
signal in which all radiometers see the same effective sky signal,
defined by the mean over all detectors, while the mean itself is not
affected by the leakage correction, since $\delta
\s^{\mathrm{leak}}_{j}$ sums to zero by construction.

These corrections are conceptually similar to those applied by the
\Planck\ LFI DPC \citep{planck2014-a03,planck2016-l02} and
\Planck\ DR4 \citep{npipe} pipelines, although implementation-wise they
differ significantly from both. Firstly, neither of the two previous
pipelines apply any beam leakage correction, while in the current work
we account for the different beam FWHMs for each radiometer
\citep{planck2014-a04} when evaluating Eq.~\eqref{eq:leak}. Secondly,
while both of the previous pipelines use a linear approximation to
the component SEDs to evaluate bandpass effects, we evaluate the full
integral in Eq.~\eqref{eq:mixmat} for each case, as described in
\citet{bp03}. Thirdly, while the DPC pipelines makes separate maps for
the raw TOD and the leakage correction, and subtract the latter as a
post-processing step, we apply the corrections directly in time-domain
prior to mapmaking, as does \Planck\ DR4. Fourthly, while the DPC
approach does not make any direct adjustments to the bandpass profiles
themselves, we fit a parametric function for each of these, as
described in the next section as part of the larger sampling
framework, and thereby achieve a better overall fit. The \Planck\ DR4
achieves a similar goal by fitting separate sky signal template terms
for the mean and derivative of the foreground SED evaluated at the
center frequency as part of their generalized mapmaker
\citep{npipe}. Finally, neither of the previous pipelines
allow for uncertainties in the astrophysical sky model itself, but
rather adopts a \commander\ sky model as a static conditional input.

Figure~\ref{fig:subleakmaps_30} shows as comparison between the
uncorrected (top row) and leakage corrected (second row) LFI 30\,GHz
map, as well as the individual contributions from bandpass leakage
corrections (third row) and beam leakage corrections (bottom
row). Comparing these maps, we immediately see that the bandpass
leakage corrections are highly significant along the Galactic plane,
with order unity corrections. Indeed, the corrections are sufficiently
large that the central Galactic plane switches sign in both Stokes $Q$
and $U$. This makes any estimates of spectral parameters
highly dependent on the leakage corrections, and joint estimation of both
leakage corrections and spectral parameters is essential; for a
specific discussion of Bayesian estimates of the spectral index of
polarized synchrotron emission, see \citet{bp14}. Likewise, the
bandpass leakage corrections at high Galactic latitudes are dominated
by the foreground monopoles, most notably those of synchrotron, AME
and free-free emission, and consistent and joint estimation of
monopoles and leakage corrections is therefore essential for
large-scale CMB extraction. A novel feature of the \BP\ processing is
component-based monopole determination, which allows more easily
self-consistent zero-level estimation across frequencies; for a
discussion of this approach, see \citet{bp13}.

Comparing the bottom two rows of Fig.~\ref{fig:subleakmaps_30}, it is
visually obvious that the bandpass corrections are generally much more
important than the beam errors on large angular scales. However, the
beam corrections can also be important for CMB power spectrum
analysis on smaller angular scales. This is illustrated in
Fig.~\ref{fig:beamleakage}, which shows a $30\deg\times30\deg$ zoom-in
of the 44\,GHz leakage correction map (top right panel) near the
Galactic South Pole. This effect is particularly strong for the
44\,GHz channel, because one of its three feeds (horn 24) has a
significantly smaller beam width ($24\arcm$ FWHM) than the other two
(horns 25 and 26; $30\arcm$ FWHM), as reported by
\citet{planck2014-a05}.

In this plot, we see that the spurious beam-induced small-scale polarization fluctuations are typically
at a level of 1--2\muK. In addition, these fluctuations are primarily
seeded by CMB temperature fluctuations, as seen by comparing the top
two panels. The in-set circle provides a visual guide that makes it
easy to identify correlations by eye. These spurious
temperature-to-polarization leakage fluctuations induce spurious
small-scale polarization modes. The bottom panel in
Fig.~\ref{fig:beamleakage} compares the angular power spectrum of the
44\,GHz channel with the \Planck\ 2018 best-fit $\Lambda$CDM $EE$
spectrum, as well as a scaled version of the $TT$ spectrum. Not only
can uncorrected beam leakage correction confuse $EE$ and $BB$
measurements, but they can also obviously be a significant contaminant
for $TE$, $TB$ and $EB$ correlations, and thereby contaminate
constraints on both standard and non-standard physics, for instance
gravitational lensing \citep[e.g.,][]{planck2015-XLI} or birefringence \citep[e.g.,][]{minami:2020}.

\subsection{Monte Carlo sampling of parametric bandpass models}
Since each radiometer bandpass is measured with
non-negligible uncertainties, our leakage correction model as defined
by Eq.~\eqref{eq:leak} is not perfect. For instance, as discussed in
Sect.~\ref{sec:newprofs} the modifications made to the 70\,GHz
bandpasses in this paper changes its overall center frequency by
0.6\,\%, and this is likely to have a non-trivial impact on foreground
estimates derived from these data, whereas relative measurement errors
between individual detectors will create bandpass leakage as discussed
above.

It is therefore important to parameterize the bandpasses themselves, fit the
free parameters to the data, and marginalize over the corresponding
uncertainties. Obviously, bandpasses have an indefinite number of degrees of
freedom, being essentially free functions of frequency, and flight data are far
from sufficient to constrain these frequency-by-frequency. It is therefore
customary to base parametric models on the laboratory measurements, $\tau_0$,
and only apply relatively mild corrections to these. For instance,
\citet{planck2014-a12} introduced a simple linear shift model,
\begin{align}
  \tau(\nu) = \tau_0(\nu+\Dbp),
  \label{eq:bpshift}
\end{align}
where $\Dbp$ denotes a linear shift in frequency space. Another
possible model is a power-law tilt model,
\begin{align}
  \tau(\nu) = \tau_0(\nu)\,\left(\frac{\nu}{\nu_\mathrm{c}}\right)^n,
  \label{eq:bpshift2}
\end{align}
where $\nu_{\mathrm{c}}$ is the center frequency, and $n$ is a
spectral index. We have implemented support for both models in our
codes, but we will focus only on the linear shift model in the
following, as we generally find that the impact of the tilt model is
too small for physically reasonable values of $n$ to have a relevant impact in terms of total map-level corrections.

\begin{figure}[t]
  \center
  \includegraphics[width=\linewidth]{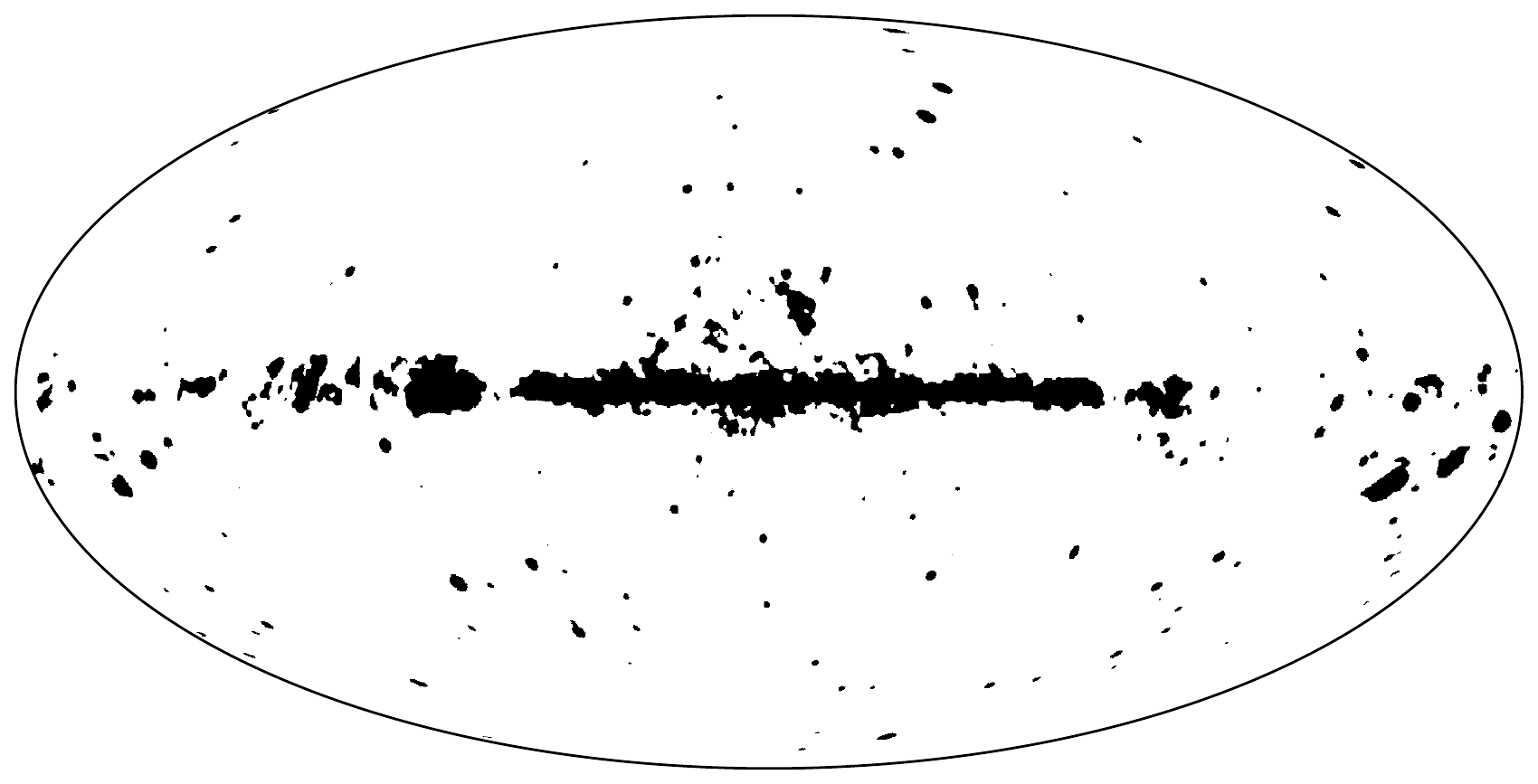}
  \caption{Mask used for bandpass correction sampling. The accepted
    sky fraction is $f_{\mathrm{sky}}=0.953$.}
  \label{fig:procmask}
\end{figure}

For the linear shift model, we split the total shift for radiometer
$j$ into two components,
\begin{equation} 
  \Delta_{\mathrm{bp}}^j = \bar{\Delta}_{\mathrm{bp}} + \delta_{\mathrm{bp}}^j.
  \label{eq:correction}
\end{equation}
Here, $\bar{\Delta}_{\mathrm{bp}}$ corresponds to an absolute
frequency shift for the overall co-added frequency band, while
$\delta_{\mathrm{bp}}^j$ is a relative frequency shift for radiometer
$j$ only, with the constraint that $\sum_j\delta_{\mathrm{bp}}^j = 0$.

Absolute and relative bandpass corrections generally have very
different impacts on the final sky maps. Intuitively, an absolute
frequency shift can be interpreted as a ``foreground-only calibration
change'', in the sense that each foreground component becomes either
weaker or brighter in the given frequency channel. The archetypal
signature of an absolute bandpass error is that the residual map
($\d_{\nu}-\s_{\nu}$) shows an imprint of the Galactic plane, but
there is no corresponding imprint of a residual CMB dipole; if there
are both a CMB dipole and a Galactic plane imprint, then the problem
is an absolute gain error. In contrast, relative bandpass errors
primarily lead to temperature-to-polarization leakage, as visualized
in Fig.~\ref{fig:subleakmaps_30}, with a pattern defined by the
Galactic intensity foregrounds modulated by the scanning strategy and
polarization angle of the experiment in question.

Based on these observations, we employ different sampling techniques for
$\bar{\Delta}_{\mathrm{bp}}$ and $\delta_{\mathrm{bp}}^j$, each with its own
likelihood and proposal density. In both cases, however, we employ a standard
Metropolis MCMC sampler with a tuned covariance matrix (see Appendix~A in
\citealp{bp01}), and we sample the absolute and relative corrections separately
and alternately in each main Gibbs step; this interleaved sampling is
exclusively for convenience of implementation, as the code becomes simpler by
considering one type of corrections at any given time.

\subsection{Absolute bandpass sampling}

To derive a likelihood for the $\bar{\Delta}_{\mathrm{bp}}$ parameter,
we note that this parameter conditionally only affects the sky signal
model through the mixing matrix in Eq.~\eqref{eq:mixmat},\footnote{In
  principle, the bandpass also affects the beam calculations, but
  since we do not apply any stochastic beam corrections in this paper,
  we ignore this effect for now.} and we may therefore form the
following data-minus-signal residual,
\begin{equation}
  \r_{\nu} = \d_{\nu}-s_{\nu}(\bar{\Delta}_{\mathrm{bp}}),
\end{equation}
where 
\begin{equation}
  \s_{\nu} = \sum_{c} \M_{cj}(\beta, \Dbp^{j})\a^c.
\end{equation}
Under the assumption that $\r_{\nu}$ may be modelled as Gaussian
noise, as per Eq.~\eqref{eq:todmodel}, this results in a
log-likelihood of the following form,
\begin{equation}
  -2\ln \mathcal{L}(\bar{\Delta}_{\mathrm{bp}}) =
  (\d_{\nu}-s_{\nu}(\bar{\Delta}_{\mathrm{bp}}))^T\N_{\nu}^{-1}(\d_{\nu}-s_{\nu}(\bar{\Delta}_{\mathrm{bp}}))
  = \chi^2(\bar{\Delta}_{\mathrm{bp}}),
  \label{eq:chisq}
\end{equation}
up to an irrelevant constant. Absolute bandpass sampling is thus
equivalent to a standard $\chi^2$ fit, and one may sample from the
corresponding conditional distribution through standard Metropolis
MCMC with an accept rate of
\begin{equation}
  a = \mathrm{min}(1,e^{-\frac{1}{2}(\chi^2_{i+1}-\chi^2_{i})}).
  \label{eq:metropolis}
\end{equation}

In practice, we apply a Galactic mask in these evaluations by setting
$N_{p}^{-1}=0$ for masked pixels. However, since the calibration signal in
question is precisely the Galactic plane, it is desirable to include
as much of the sky as possible, and we only exclude the central parts
of the Milky Way (within which foreground modelling is very
complicated), as well as particularly bright points sources, such as
Tau-A. The actual mask used is shown as a black region in
Fig.~\ref{fig:procmask}, and excludes 4.7\,\% of the sky.

Finally, following \citet{planck2014-a12}, we will only estimate an
absolute correction for the LFI 30\,GHz channel. Allowing all channel
corrections to be fitted freely is impossible, as this would result in
perfect degeneracies between the bandpass parameters and the
foreground SED parameters. The motivation for fitting the 30\,GHz
channel is simply that this parameter turns out to be sufficiently
non-degenerate to allow a robust fit, while at the same time
non-negligible residuals occur when it is not fitted. The 30\,GHz
bandpass parameter is in practice constrained by its
nearest \WMAP\ channel, namely the Ka-band at 33\,GHz.

\subsection{Relative bandpass sampling}

For the relative bandpass corrections, $\delta_{\mathrm{bp}}^j$, we adopt an
alternative likelihood that is inspired by the spurious map approach pioneered
by \citet{page2007} for CMB polarization purposes. The motivation for this is
that relative bandpass corrections primarily have a real-world impact for
polarization, while a full and formally statistically correct $\chi^2$ fit, as
defined by Eq.~\eqref{eq:chisq}, is vastly dominated by the high signal-to-noise
ratio of the intensity signal. In practice, a global fit would therefore tend to
use the extra degrees of freedom to fit intensity foreground SED errors, rather
than large-scale polarization artefacts.

As noted by \citet{page2007}, relative bandpass errors cause
intensity-to-polarization leakage with a very specific observational
signature, namely that the spurious polarization signal does not depend on the
polarization angle orientation of a given detector, but only on its
bandpass properties. They use this to define an additional correction
map for each radiometer, $S_j$, that corresponds to the difference
between the intensity signal seen by detector $j$ and the mean over
all radiometers. Thus, the time-domain signal measured by detector $j$
may be written in the form
\begin{equation}
    s_{j} = T + Q\cos2\psi_j + U\sin2\psi_j +
    \sum_{i=1}^{N_{\mathrm{det}}-1}S_i \delta_{ij},
\end{equation}
and the pointing matrix $\P$ in Eq.~\eqref{eq:todmodel} may be modified
accordingly.

  \begin{figure}[t] 
    \center
    \includegraphics[width=0.95\linewidth]{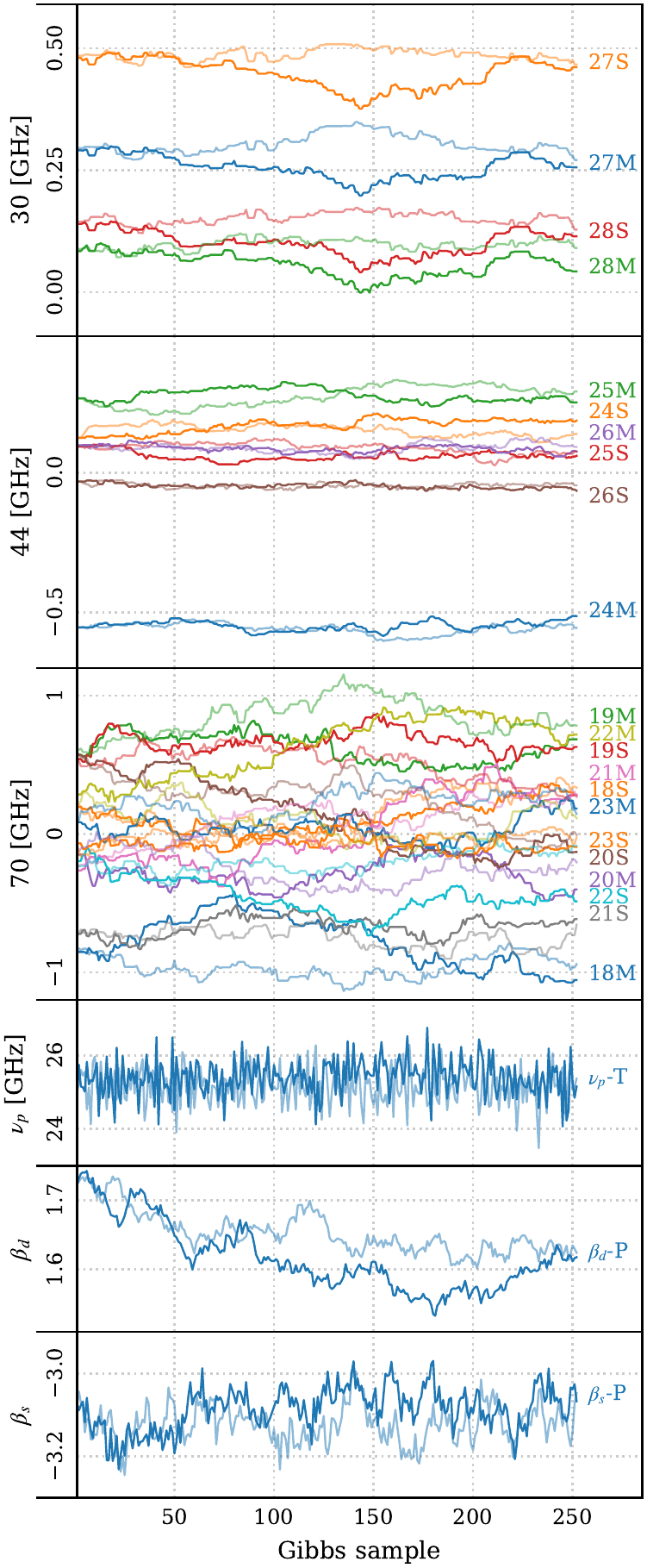}
    \caption{Markov chain trace plots for bandpass and spectral index parameters. The two chains are indicated with different opacities.}
    \label{fig:traceplot}
  \end{figure}

In general, one can only solve for $N-1$ spurious sky signals, where
$N$ is the number of radiometers in a band, to avoid a perfect
degeneracy with the mean intensity signal. For simplicity, let us
therefore consider a minimal case with $N=2$. In this case, the
single-pixel mapmaking equation reads
{\fontsize{7}{4}\selectfont
\begin{equation}
    \left[\begin{array}{cccc}
        1 & \cos 2\psi & \sin 2\psi & \,\delta_{1j} \\
        \cos 2\psi & \cos^2 2\psi & \cos 2\psi \sin 2\psi & \cos 2\psi\,
        \,\delta_{1j} \\
        \sin 2\psi & \sin 2\psi\cos 2\psi & \sin^2 2\psi & \sin 2\psi\,
        \,\delta_{1j} \\
        \,\delta_{1j} & \cos 2\psi\,\delta_{1i} & \sin 2\psi\,\delta_{1j} & \,\delta_{1j} \\
    \end{array}\right]
    \left[\begin{array}{c}
        T \\ Q \\ U \\ S_1
    \end{array}\right]
    =
    \left[\begin{array}{c}
        d \\ d\cos 2\psi  \\ d\sin 2\psi  \\ d\,\delta_{1j}
    \end{array}\right].
    \label{eq:Smap}
\end{equation}
}\normalfont

For the tightly interconnected \WMAP\ scanning strategy, this
equation may be solved pixel-by-pixel without inducing a prohibitive
noise penalty, and, as a result, \citet{page2007} simply chose to
deliver polarization sky maps that are explicitly marginalized over
$S$. However, this is not possible for the \Planck\ scanning strategy,
for which the polarization angle of a given detector only varies by a
few tens of degrees over large areas of the sky
\citep{planck2013-p01}. For these, the condition number of the matrix
in Eq.~\eqref{eq:Smap} leads to a massive noise increase, to the point
that the map becomes unusable for astrophysical and cosmological
analysis.

However, even though the noise per pixel is excessive, the aggregated
signal-to-noise ratio in $S_j$ across the full sky is still high. In this
paper, we therefore instead propose to use the spurious map approach
to fit the small number of relative bandpass shifts through the
following goodness-of-fit quantity,
\begin{equation}
  \chi^2 = \sum_{j=1}^{N_{\mathrm{det}}-1} \sum_p \left(\frac{S_j(p)}{\sigma_{j}(p)}\right)^2.
\end{equation}
Here, $\sigma_j(p)$, is the uncertainty arising from the $IQUS$
solution above, which is defined as the diagonal element of the
inverse coupling matrix in Eq.~\eqref{eq:Smap}. The rest of the
algorithm is identical to that described in the previous section, just
with a different $\chi^2$ expression in Eq.~\eqref{eq:metropolis}.

Intuitively, this approach combines the \Planck\ idea of using a
parametric foreground model to perform relative bandpass corrections
with the spurious map approach from \WMAP, to optimize the
free parameters using only polarization information. As such, the
algorithm significantly reduces temperature-to-polarization
leakage through a very small number of additional degrees of freedom
and a negligible noise boost.

\section{Results}
\label{sec:results}

  \begin{figure}[t] 
    \center
    \includegraphics[width=0.4892\linewidth]{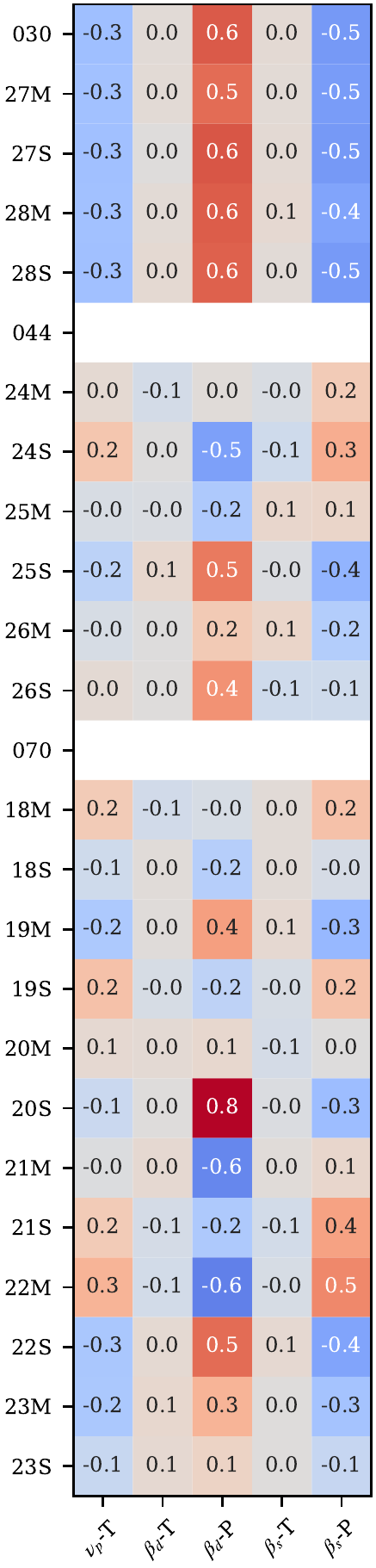}
    \caption{Pearson's correlations between bandpass correction
    and other parameters in the global \BP\ model as evaluated
    directly from Markov chain.} 
    \label{fig:correlations}
  \end{figure}

\begin{figure}
  \begin{minipage}{\linewidth}
    \centering
    \newdimen\tblskip \tblskip=5pt
    \captionof{table}{Bandpass correction posterior means and standard
      deviations for all \Planck\ LFI frequency bands and
      individual radiometers. Note that the absolute bandpass
      corrections at 44 and 70\,GHz are fixed to zero. \label{tab:corrections}}
    \vskip -4mm
    \footnotesize
    \setbox\tablebox=\vbox{
      \newdimen\digitwidth
      \setbox0=\hbox{\rm 0}
      \digitwidth=\wd0
      \catcode`*=\active
      \def*{\kern\digitwidth}
      \newdimen\dpwidth
      \setbox0=\hbox{.}
      \dpwidth=\wd0
      \catcode`!=\active
      \def!{\kern\dpwidth}
      \halign{\hbox to 4cm{#\leaderfil}\tabskip 1.5em&
        \hfil$#$\hfil \tabskip 1.em&
        #\hfil \tabskip 0em\cr
    \noalign{\doubleline}
    \omit\sc Radiometer \hfil& \Dbp\, [\GHz]  \cr
    \noalign{\vskip 3pt\hrule\vskip 5pt}
    \omit\textbf{LFI 30\,GHz}\hfil     & *0.24\pm0.03 \cr
    \noalign{\vskip 2pt}
    \hskip 15pt 27M         & *0.28\pm0.03 \cr
    \hskip 15pt 27S         & *0.47\pm0.03 \cr
    \hskip 15pt 28M         & *0.08\pm0.03 \cr
    \hskip 15pt 28S         & *0.13\pm0.03 \cr
    \noalign{\vskip 5pt}
    \omit\textbf{LFI 44\,GHz}\hfil   & \cdots \cr
    \noalign{\vskip 2pt}
    \hskip 15pt 24M        & -0.55\pm0.02 \cr
    \hskip 15pt 24S        & *0.16\pm0.02 \cr
    \hskip 15pt 25M        & *0.28\pm0.03 \cr
    \hskip 15pt 25S        & *0.07\pm0.02 \cr
    \hskip 15pt 26M        & *0.09\pm0.01 \cr
    \hskip 15pt 26S        & -0.05\pm0.01 \cr
    \noalign{\vskip 5pt}
    \omit\textbf{LFI 70\,GHz}\hfil  & \cdots\cr
    \noalign{\vskip 2pt}
    \hskip 15pt 18M        & -0.87\pm0.17\cr
    \hskip 15pt 18S        & *0.14\pm0.12\cr
    \hskip 15pt 19M        & *0.73\pm0.16\cr
    \hskip 15pt 19S        & *0.59\pm0.12\cr
    \hskip 15pt 20M        & -0.27\pm0.12\cr
    \hskip 15pt 20S        & *0.20\pm0.22\cr
    \hskip 15pt 21M        & *0.06\pm0.20\cr
    \hskip 15pt 21S        & -0.71\pm0.08\cr
    \hskip 15pt 22M        & *0.36\pm0.32\cr
    \hskip 15pt 22S        & -0.32\pm0.16\cr
    \hskip 15pt 23M        & *0.13\pm0.13\cr
    \hskip 15pt 23S        & -0.05\pm0.05\cr
    \noalign{\vskip 5pt\hrule\vskip 5pt}}}
    \endPlancktable
    \end{minipage}
  \end{figure}

  \begin{figure} 
    \center
    \includegraphics[width=\linewidth]{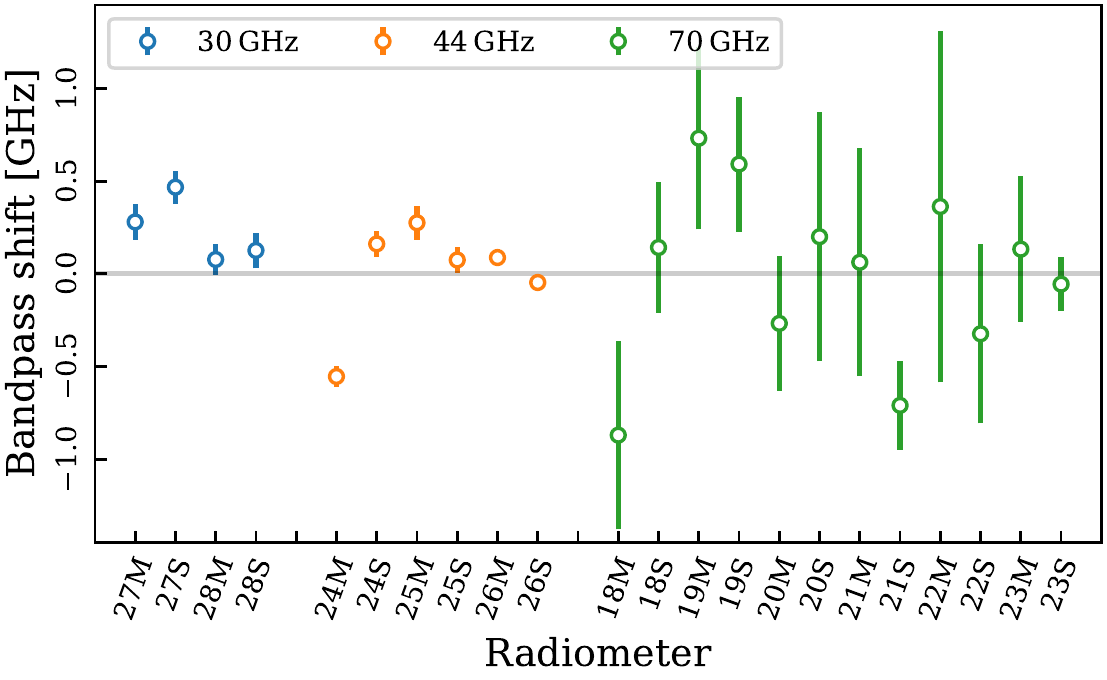}
    \caption{Estimated bandpass corrections for each LFI radiometer. Error bars indicate $\pm3\sigma$ uncertainties. These parameters are constrained to have vanishing mean within each frequency (except for 30\,GHz, which has an equal absolute shift to all radiometers as well), and are as such strongly correlated.}
    \label{fig:bpshiftfinal}
  \end{figure}
  
  \begin{figure*}[p] 
    \center
    \begin{subfigure}[t]{\linewidth}
      \includegraphics[width=0.32\linewidth]{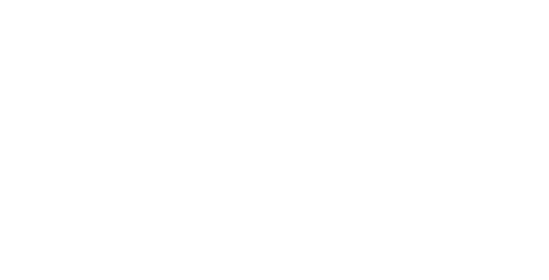}
      \includegraphics[width=0.32\linewidth]{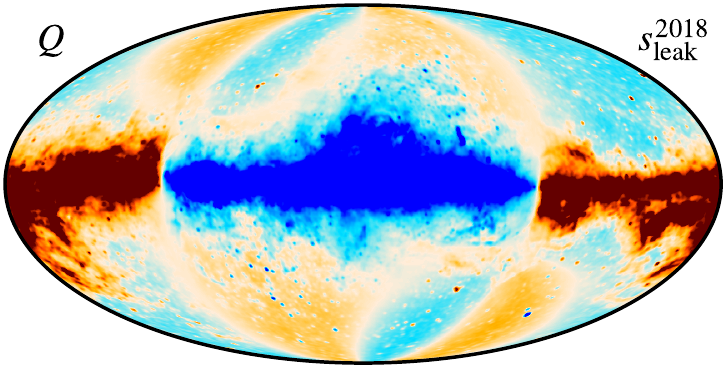}
      \includegraphics[width=0.32\linewidth]{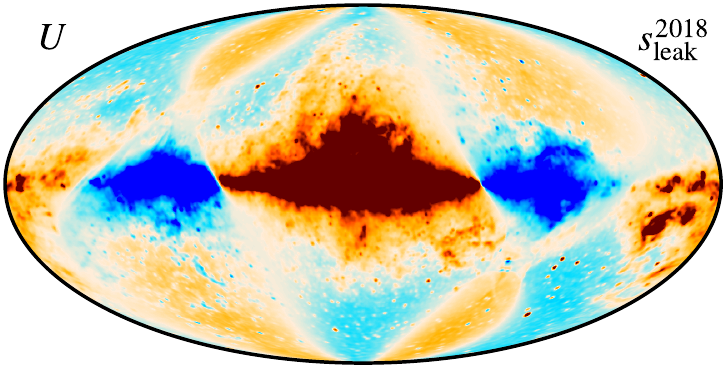}\\
      \includegraphics[width=0.32\linewidth]{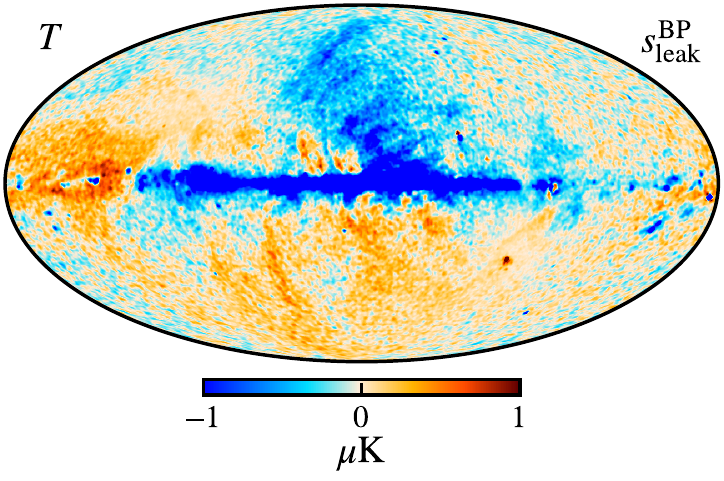}
      \includegraphics[width=0.32\linewidth]{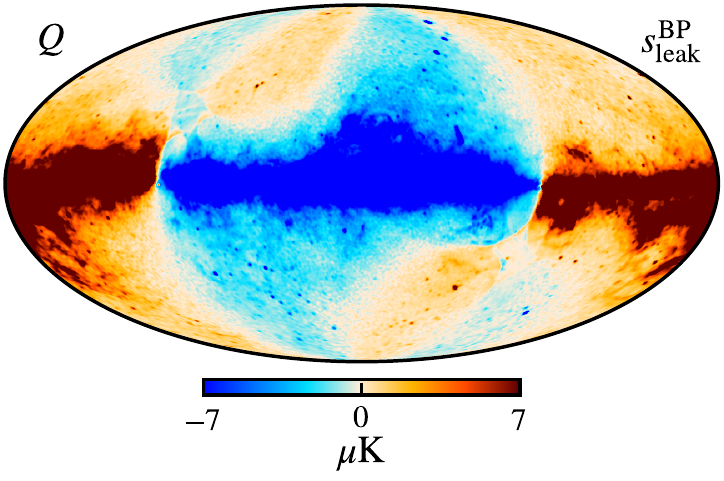}
      \includegraphics[width=0.32\linewidth]{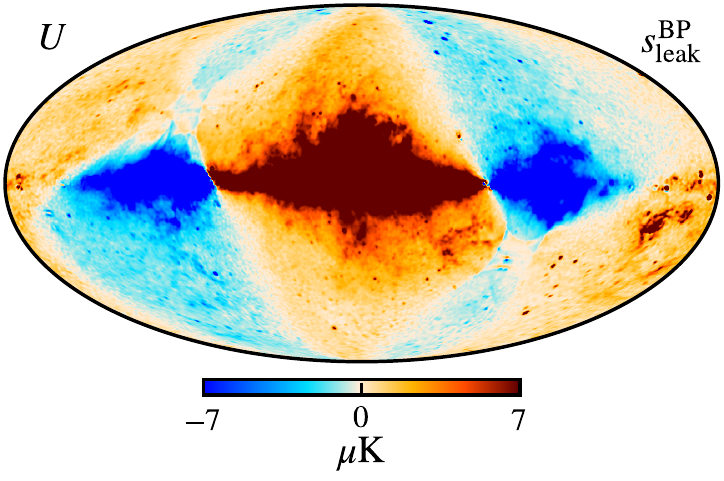}
      \caption{30\,GHz}
      \vspace*{3mm}
    \end{subfigure}
    \begin{subfigure}[t]{\linewidth}
      \includegraphics[width=0.32\linewidth]{figs/whitemap.pdf}
      \includegraphics[width=0.32\linewidth]{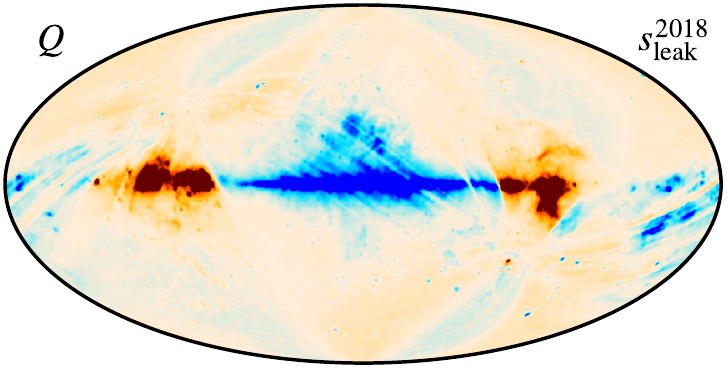}
      \includegraphics[width=0.32\linewidth]{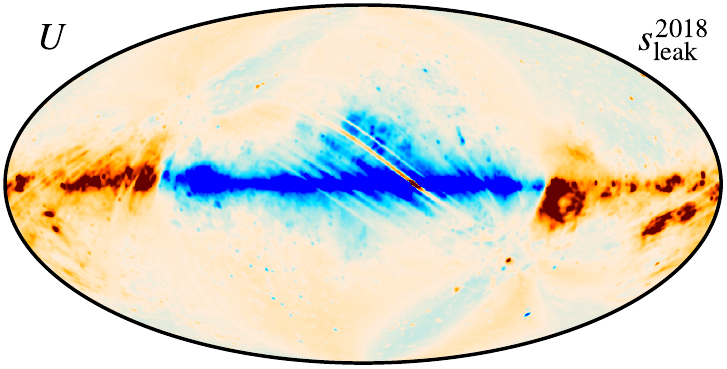}\\
      \includegraphics[width=0.32\linewidth]{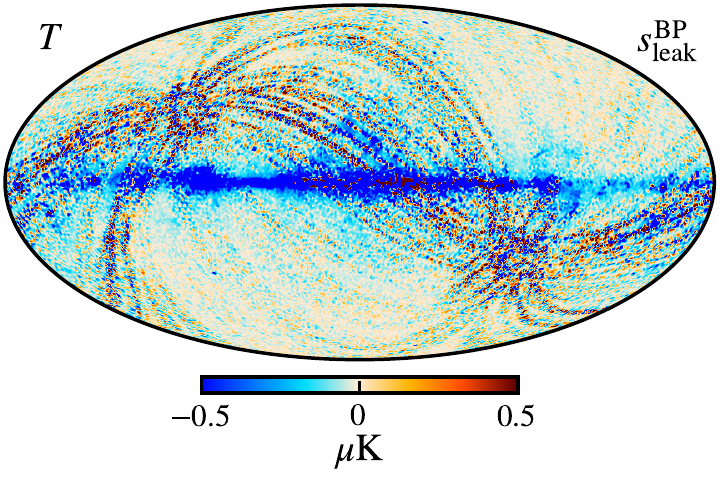}
      \includegraphics[width=0.32\linewidth]{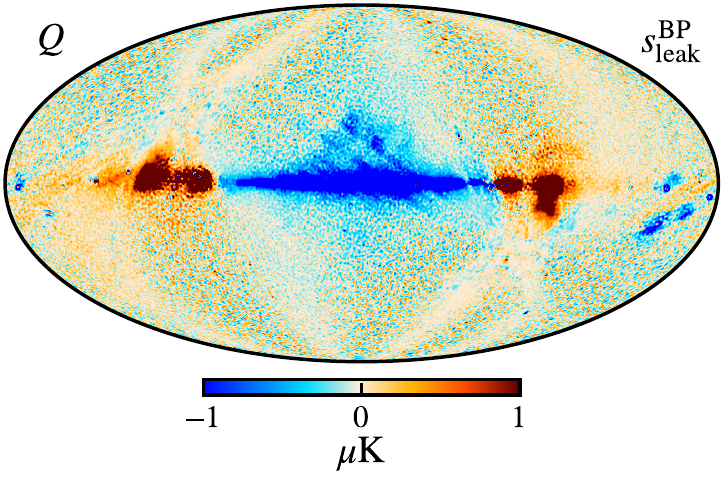}
      \includegraphics[width=0.32\linewidth]{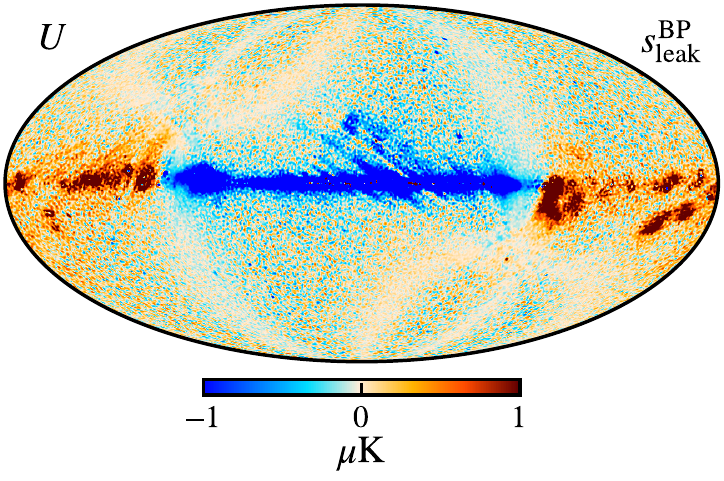}
      \caption{44\,GHz}
      \vspace*{3mm}
    \end{subfigure}
    \begin{subfigure}[t]{\linewidth}
      \includegraphics[width=0.32\linewidth]{figs/whitemap.pdf}
      \includegraphics[width=0.32\linewidth]{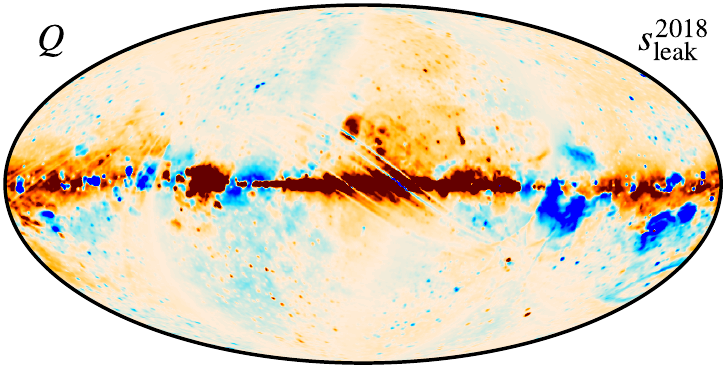}
      \includegraphics[width=0.32\linewidth]{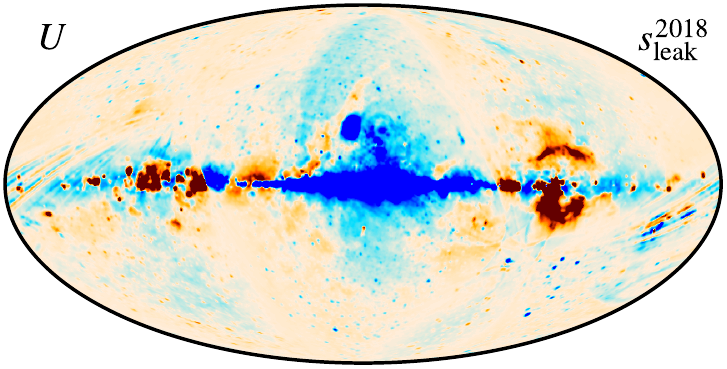}\\
      \includegraphics[width=0.32\linewidth]{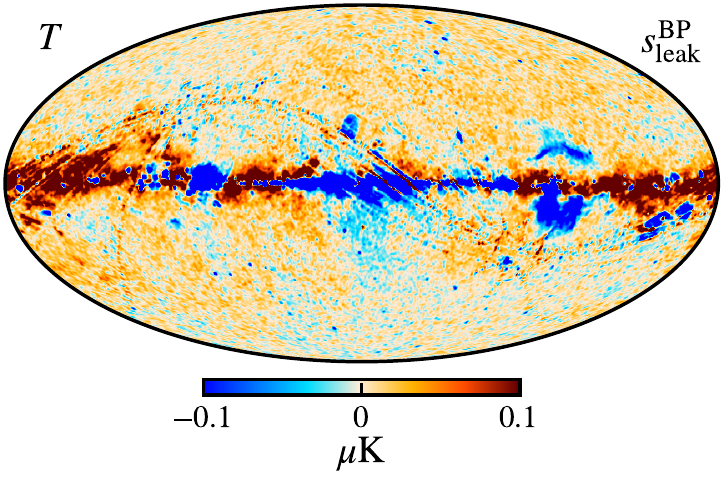}
      \includegraphics[width=0.32\linewidth]{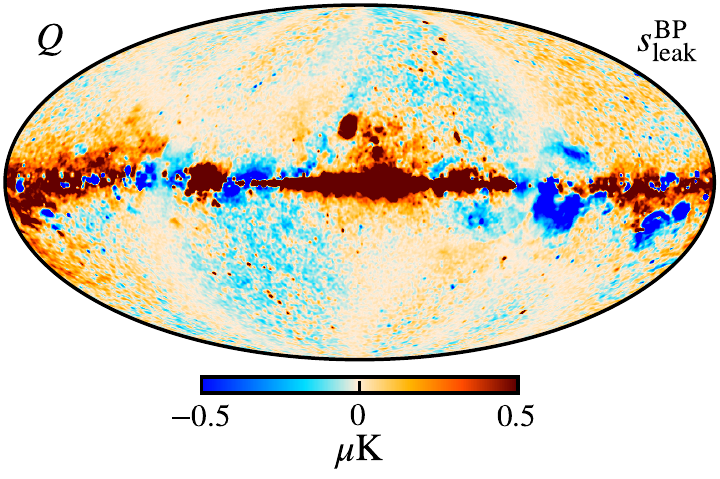}
      \includegraphics[width=0.32\linewidth]{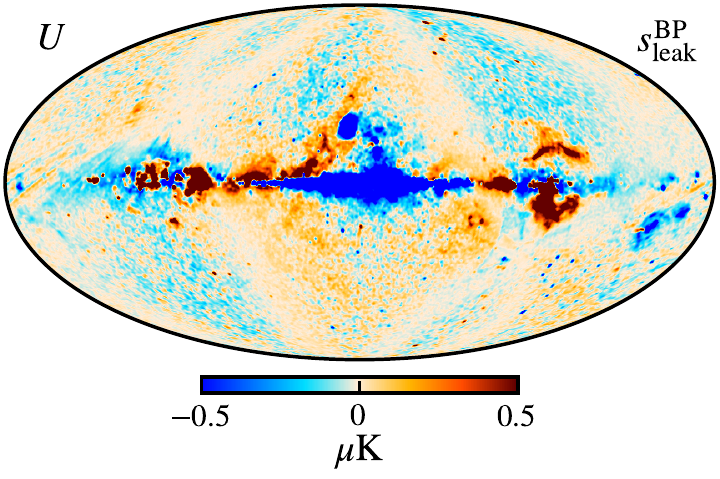}
      \caption{70\,GHz}
    \end{subfigure}
    \caption{Full leakage correction maps for the 30\,GHz (a), 44\,GHz (b) and 70\,GHz (c) bands in Stokes $T$, $Q$, and $U$ from the official \Planck\ 2018 processing (top) and \BP\ (bottom). All maps have been smoothed to an angular resolution of 1\deg\ FWHM.}\label{fig:leakmaps}
  \end{figure*}
  
  \begin{figure*}[t] 
    \center
    \includegraphics[width=0.32\linewidth]{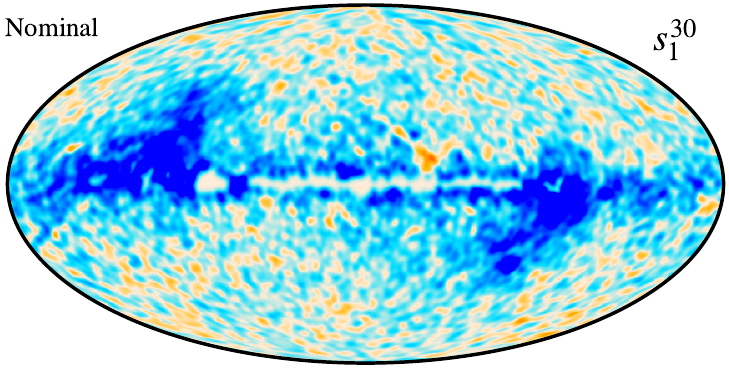}
    \includegraphics[width=0.32\linewidth]{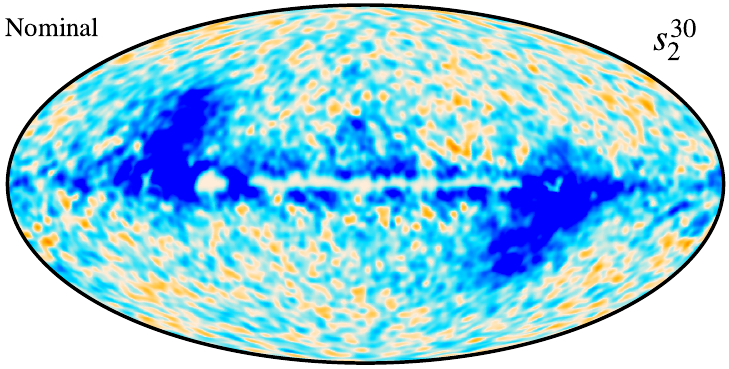}
    \includegraphics[width=0.32\linewidth]{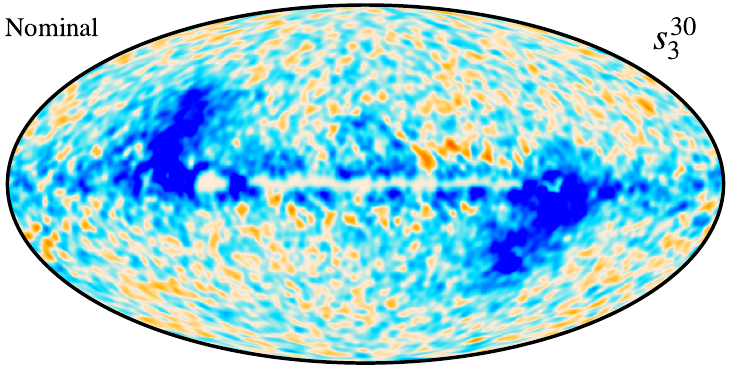}\\
    \includegraphics[width=0.32\linewidth]{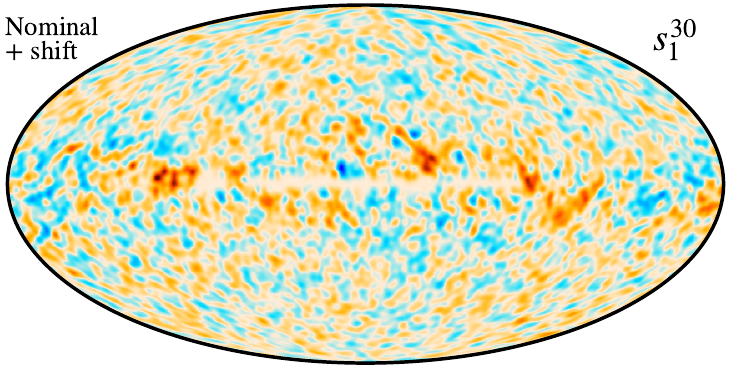}
    \includegraphics[width=0.32\linewidth]{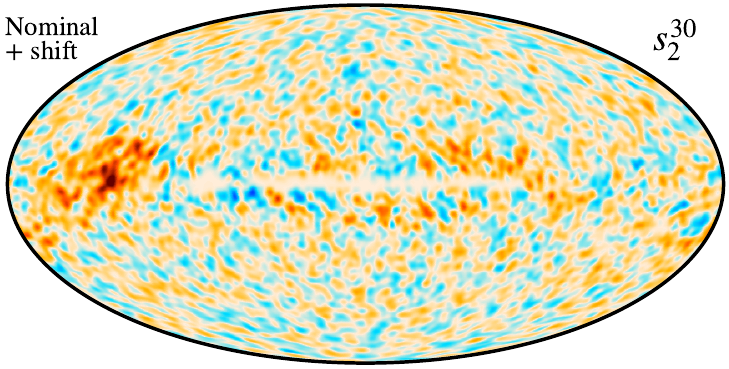}
    \includegraphics[width=0.32\linewidth]{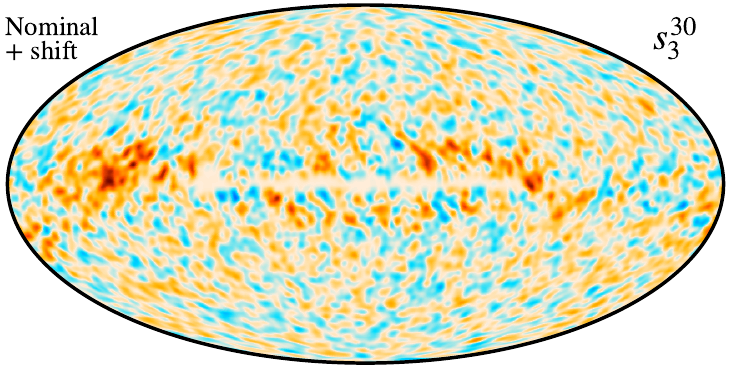}\\
    \includegraphics[width=0.32\linewidth]{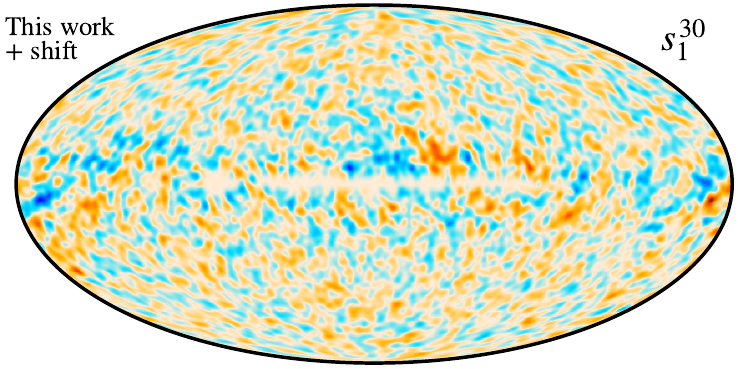}
    \includegraphics[width=0.32\linewidth]{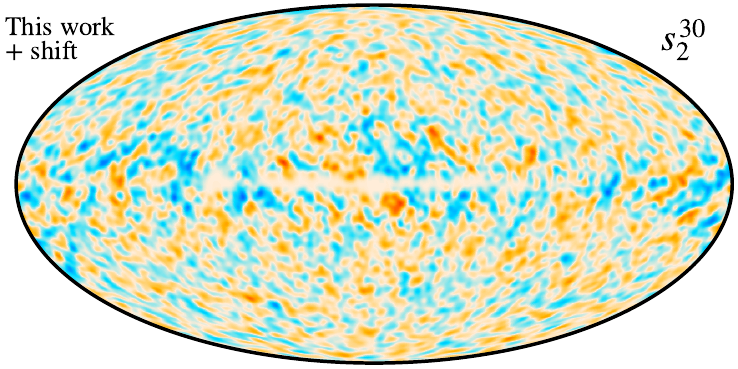}
    \includegraphics[width=0.32\linewidth]{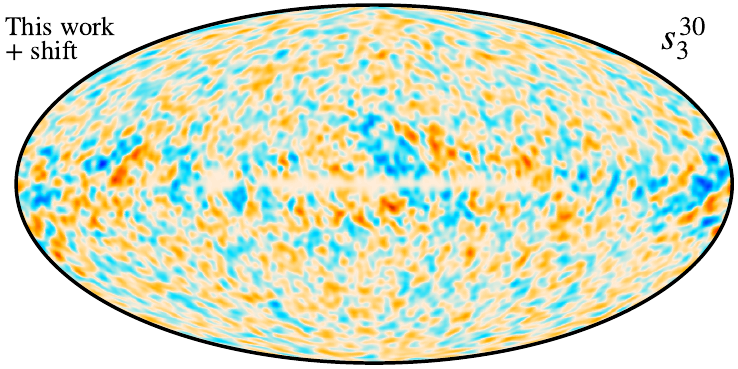}\\
    \vspace{0.2cm}
    \includegraphics[width=0.25\linewidth]{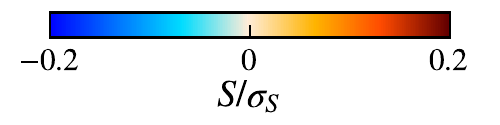}
    \caption{Spurious maps for the 30\,GHz band with nominal profiles without corrections (\textit{top}), nominal profiles with corrections (\textit{middle}) and new profiles with corrections (\textit{bottom}) using the processing mask showed in Fig.~\ref{fig:procmask} ($f_{\mathrm{sky}}=0.953$), and smoothed with a gaussian beam of 3$\deg$ FWHM. This quantity is unitless, as it is defined as $S/\sigma_{S}$.}
    \label{fig:smapdiffs} 
  \end{figure*}

\subsection{Posterior summary}

We are now finally ready to present the results from the above algorithms as
applied within the \BP\ analysis framework, and we start by inspecting the
resulting Markov chains and their internal correlations. As described by
\citet{bp01}, two independent Monte Carlo chains are produced in the main \BP\
processing, each resulting in 750 samples. Figure~\ref{fig:traceplot} shows a
representative subset of these, where in addition to the individual detector
bandpass shifts, we also include the AME $\nu_p$ parameter in temperature, as
well as the spectral index in polarization for both thermal dust and
synchrotron, $\beta_d$ and $\beta_s$, respectively. We note that only the two
sampled sky-regions of the synchrotron index are included and not those that are
drawn from a prior distribution, see \citet{bp14} for a full overview. 

Overall, we see that the correlation length is substantial in these
chains, and it is clear that several of the Metropolis step lengths would
benefit from further optimization in a future run. Indeed, using the current
chains to tune a Metropolis proposal matrix for a future iteration of the
analysis appears to be promising. However, there is in general no doubt that bandpass
corrections are among the most difficult parameters to sample in the entire \BP\
data model, because bandpass corrections are global, and do not only affect
leakage corrections, but also unit conversions, and thereby the entire
foreground model. This global impact leads to a long Markov chain correlation
length.

In Fig.~\ref{fig:correlations}, we plot Pearson's correlation
coefficient, $p$, for various parameter pairs. Overall, we see that
the bandpass parameters are most strongly correlated with the AME peak
frequency, $\nu_{\mathrm{p}}$ in intensity, and the synchrotron and
thermal dust spectral indices, $\beta_{\mathrm{s}}$ and
$\beta_{\mathrm{d}}$, in polarization, for which correlations around
$|p|\approx0.5$ are observed. An in-depth discussion on temperature
foreground degeneracies can be found in \citet{bp13}. We also note
that this plot serves as a powerful reminder of the usefulness of
global cross-experiment analysis, as improved constraints on either
$\nu_{\mathrm{p}}$, $\beta_{\mathrm{s}}$ or $\beta_{\mathrm{d}}$ from
for instance \Planck\ HFI \citep{planck2016-l03}, C-BASS
\citep{king2010} or QUIJOTE \citep{QUIJOTE_I_2015} will translate
directly into improved constraints on the \Planck\ LFI bandpasses, and
therefore better maps overall. Generally speaking, older data sets may
almost always be improved when new experiments become available.

The average properties of the Markov chains are listed in
Table~\ref{tab:corrections}, with the posterior mean and standard
deviation for each bandpass correction parameter, both for co-added
frequency channels and individual radiometers, and the same
information is visualized in Fig.~\ref{fig:bpshiftfinal}.

Figure~\ref{fig:leakmaps} compares the \BP\ posterior mean leakage
maps with the corresponding \Planck\ 2018 LFI leakage maps for all
three LFI channels. As discussed in Sect.~\ref{sec:leakcorr}, there
are several algorithmic differences between these two
pipelines. Firstly, the \Planck\ 2018 approach does not account for beam
mismatch leakage, and, thus, we see much less small-scale fluctuations
in these maps compared to \BP; this is particularly striking in the
44\,GHz channel.
Secondly, we note that the \Planck\ 2018 correction maps exhibit
significantly weaker large-scale corrections at high Galactic
latitudes compared to \BP. We have traced this effect down to a
difference in the net foreground monopole in the two approaches, and,
in particular, we have found that the \BP\ 30\,GHz leakage map takes a
very similar shape if we add an artificial negative monopole of
$-200\muK$ to the AME amplitude map. In this respect, we note that the
\Planck\ 2018 pipeline uses the \texttt{Madam} map-making code
\citep{keihanen2005}, which explicitly removes all monopole terms from
the co-added map \citep{planck2016-l02}. At the same time, the
\commander-based foreground maps used to generate the correction
templates do include a physical estimate of the monopoles. To account
for this issue, the DPC processing generated a new set of correction
maps using the \Planck\ \texttt{levelS} simulation package
\citep{reinecke2006}, creating TODs from the \commander\ leakage map
and correction factors, and binned these into maps using the same
\texttt{Madam} map-making code, resulting in correction maps without
the monopole term, from which the final correction templates were
generated. In retrospect, it appears that this rather involved
pipeline did underestimate the monopole contribution in the final
maps, and this serves as a useful reminder of an important benefit of
an integrated end-to-end pipeline: Passing data objects from one
operation to the next in a self-consistent manner becomes much more
transparent when all parts of the code use the same data model. 

\subsection{Comparison of nominal and corrected LFI bandpasses}

  \begin{figure}
    \center
    \includegraphics[width=\linewidth]{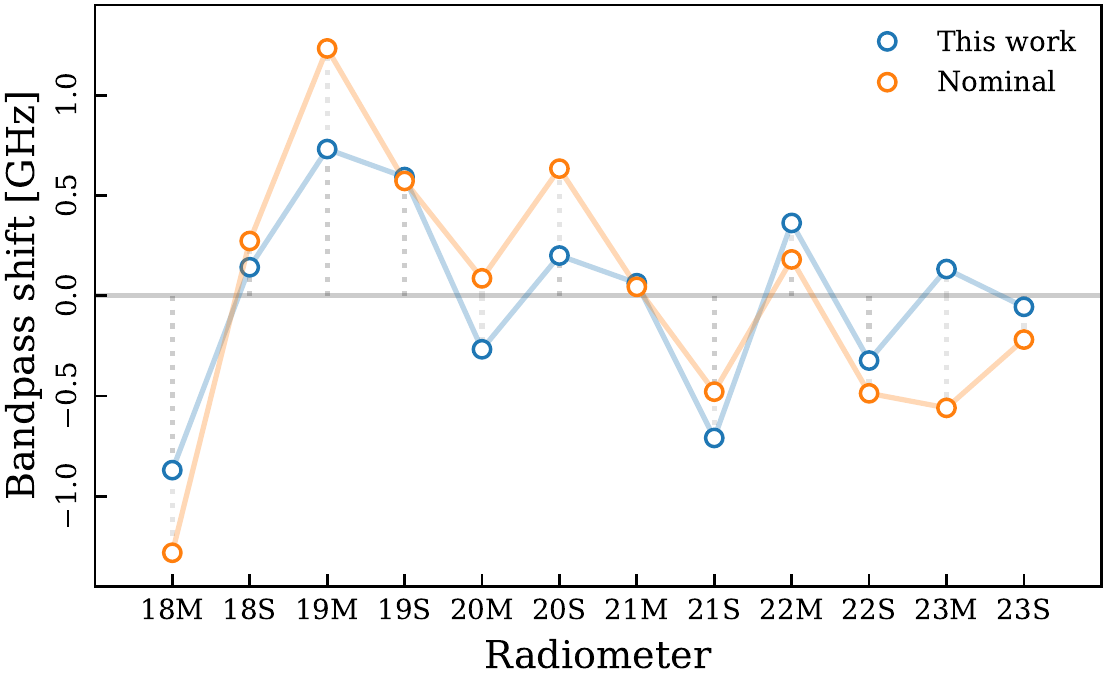}
    \caption{Estimated bandpass corrections for the 70\,GHz radiometer. Estimated bandpass corrections for each LFI radiometer. These parameters are constrained to have vanishing mean, and are as such strongly correlated.}
    \label{fig:bpshift70}
  \end{figure}

As discussed in Sect.~\ref{sec:newprofs}, we have applied a number of
pre-processing steps to the \Planck\ LFI bandpasses before performing
the main \BP\ end-to-end analysis. In this section, we consider the
impact of these changes in terms of some key data products that
highlight their effects.

The first quantity we consider is the spurious map, $S_j$, as defined by
Eq.~\eqref{eq:Smap}. These are shown for the 30\,GHz channel in
Fig.~\ref{fig:smapdiffs} for three different analysis configurations, all in the
form of $S_j/\sigma_j$, masked by the bandpass processing mask, and smoothed to
$3\deg$ FWHM. The top row shows the three 30\,GHz spurious maps that result if
one attempts to produce maps with no bandpass corrections at all, and simply
takes the publicly available profiles at face value. Here we see coherent
structures that are clearly inconsistent with noise, and morphologically
associated with the Galactic plane and the so-called ``\Planck\ Deep Fields''
(centered on the Ecliptic poles). In these regions, the $S$ signal-to-noise
ratio is vastly higher than in the Ecliptic plane, because of the dense
polarization angle sampling of the \Planck\ scanning strategy. The second row
shows a similar case, using the nominal LFI bandpasses, but this time allowing a
frequency shift per radiometer, as described in Sect.~\ref{sec:methods}. The
residuals are clearly reduced, indicating that the fitting algorithm does work
as expected. Finally, the bottom row shows the same, but now using the corrected
LFI bandpasses from Sect.~\ref{sec:newprofs} as input. At this point, the
residuals are significantly closer to randomly distributed, with only small
residuals appearing near the Galactic center.

Figure~\ref{fig:bpshift70} shows a comparison of posterior mean and
standard deviations for the 70\,GHz bandpass corrections when using
the nominal (orange points) and corrected (blue points)
profiles. Here, we see that the nominal profiles generally require
much larger correction factors than the corrected profiles. In
particular, 18M sees $\Dbp\sim1.3$\,GHz using the nominal profiles as
opposed to $\sim0.8\,$GHz for the new. This is not surprising,
considering that this profile exhibits some of the strongest
systematic artefacts in Fig.~\ref{fig:bp_LFI}. 

\section{Conclusions}
\label{sec:conclusion}

In this paper, we have discussed how bandpass and beam leakage effects may be
mitigated in a Bayesian CMB analysis pipeline, and summarized the results from
the \BP\ pipeline. As a preparatory step, we have also provided a set of
corrected LFI bandpass profiles that correct for known systematic features
dating back to the ground calibration phase and previously reported in the
literature. The most notable result from these corrections is that the overall
70\,GHz center frequency is shifted up by 0.6\,\%, and simple map-based
comparisons show that differences in the resulting frequency maps are on the
order of a couple of microkelvins. 

We argue that the proposed algorithms are substantially simpler to implement
compared to previous methods, due to the tight integration between astrophysical
component separation and low-level mapmaking; all the required components are
already available from different parts of the \BP\ framework. As a result, the
leakage correction may actually be defined in terms of two or three very simple
equations, and the practical code implementation amounts to a few hundred lines
of code. We also note that beam leakage corrections are trivial to implement,
simply by accounting for the different detector beam responses when scanning the
model sky with the pointing operator.

We believe that the importance of these methods will become increasingly
critical for next-generation experiments. While the current \Planck\ LFI
polarization observations are intrinsically noise dominated, and the various
corrections discussed in this paper are relatively minor compared to the overall
noise level, the same will not hold true for future $B$-mode experiments such as
LiteBIRD or CMB-S4; for these, establishing highly accurate bandpass and beam
leakage corrections will be absolutely essential in order to reach the required
nanokelvin accuracy.

\begin{acknowledgements}
  We thank Prof.\ Pedro Ferreira and Dr.\ Charles Lawrence for useful suggestions, comments and 
  discussions. We also thank the entire \Planck\ and \WMAP\ teams for
  invaluable support and discussions, and for their dedicated efforts
  through several decades without which this work would not be
  possible. The current work has received funding from the European
  Union’s Horizon 2020 research and innovation programme under grant
  agreement numbers 776282 (COMPET-4; \BP), 772253 (ERC;
  \textsc{bits2cosmology}), and 819478 (ERC; \textsc{Cosmoglobe}). In
  addition, the collaboration acknowledges support from ESA; ASI and
  INAF (Italy); NASA and DoE (USA); Tekes, Academy of Finland (grant
   no.\ 295113), CSC, and Magnus Ehrnrooth foundation (Finland); RCN
  (Norway; grant nos.\ 263011, 274990); and PRACE (EU).
\end{acknowledgements}

\bibliographystyle{aa}

\bibliography{Planck_bib,BP_bibliography}

\begin{thebibliography}{34}
\expandafter\ifx\csname natexlab\endcsname\relax\def\natexlab#1{#1}\fi

\bibitem[{{Ade} {et~al.}(2021){Ade}, {Ahmed}, {Amiri}, {Barkats}, {Thakur},
  {Bischoff}, {Beck}, {Bock}, {Boenish}, {Bullock}, {Buza}, {Cheshire},
  {Connors}, {Cornelison}, {Crumrine}, {Cukierman}, {Denison}, {Dierickx},
  {Duband}, {Eiben}, {Fatigoni}, {Filippini}, {Fliescher}, {Goeckner-Wald},
  {Goldfinger}, {Grayson}, {Grimes}, {Hall}, {Halal}, {Halpern}, {Hand},
  {Harrison}, {Henderson}, {Hildebrandt}, {Hilton}, {Hubmayr}, {Hui}, {Irwin},
  {Kang}, {Karkare}, {Karpel}, {Kefeli}, {Kernasovskiy}, {Kovac}, {Kuo}, {Lau},
  {Leitch}, {Lennox}, {Megerian}, {Minutolo}, {Moncelsi}, {Nakato}, {Namikawa},
  {Nguyen}, {O'Brient}, {Ogburn}, {Palladino}, {Prouve}, {Pryke}, {Racine},
  {Reintsema}, {Richter}, {Schillaci}, {Schwarz}, {Schmitt}, {Sheehy},
  {Soliman}, {Germaine}, {Steinbach}, {Sudiwala}, {Teply}, {Thompson}, {Tolan},
  {Tucker}, {Turner}, {Umilt{\`a}}, {Verg{\`e}s}, {Vieregg}, {Wandui}, {Weber},
  {Wiebe}, {Willmert}, {Wong}, {Wu}, {Yang}, {Yoon}, {Young}, {Yu}, {Zeng},
  {Zhang}, {Zhang}, \& {Bicep/Keck Collaboration}}]{bicep2021}
{Ade}, P.~A.~R., {Ahmed}, Z., {Amiri}, M., {et~al.} 2021, \prl, 127, 151301

\bibitem[{{Andersen et al.}(2022)}]{bp13}
{Andersen et al.} 2022, \aap, in preparation [\eprint[arXiv]{201x.xxxxx}]

\bibitem[{{BeyondPlanck}(2022)}]{bp01}
{BeyondPlanck}. 2022, \aap, in preparation [\eprint[arXiv]{2011.05609}]

\bibitem[{{Colombo et al.}(2022)}]{bp11}
{Colombo et al.} 2022, \aap, in preparation [\eprint[arXiv]{201x.xxxxx}]

\bibitem[{{Delouis} {et~al.}(2019){Delouis}, {Pagano}, {Mottet}, {Puget}, \&
  {Vibert}}]{delouis:2019}
{Delouis}, J.~M., {Pagano}, L., {Mottet}, S., {Puget}, J.~L., \& {Vibert}, L.
  2019, \aap, 629, A38

\bibitem[{{Eriksen} {et~al.}(2008){Eriksen}, {Jewell}, {Dickinson}, {Banday},
  {G{\'o}rski}, \& {Lawrence}}]{eriksen2008}
{Eriksen}, H.~K., {Jewell}, J.~B., {Dickinson}, C., {et~al.} 2008, \apj, 676,
  10

\bibitem[{{Eriksen} {et~al.}(2004){Eriksen}, {O'Dwyer}, {Jewell}, {Wand elt},
  {Larson}, {G{\'o}rski}, {Levin}, {Banday}, \& {Lilje}}]{eriksen:2004}
{Eriksen}, H.~K., {O'Dwyer}, I.~J., {Jewell}, J.~B., {et~al.} 2004, \apjs, 155,
  227

\bibitem[{{Galloway et al.}(2022)}]{bp03}
{Galloway et al.} 2022, \aap, in preparation [\eprint[arXiv]{201x.xxxxx}]

\bibitem[{Geman \& Geman(1984)}]{geman:1984}
Geman, S. \& Geman, D. 1984, IEEE Trans. Pattern Anal. Mach. Intell., 6, 721

\bibitem[{{G{\'e}nova-Santos} {et~al.}(2015){G{\'e}nova-Santos},
  {Rubi{\~n}o-Mart{\'\i}n}, {Rebolo}, {Pel{\'a}ez-Santos},
  {L{\'o}pez-Caraballo}, {Harper}, {Watson}, {Ashdown}, {Barreiro},
  {Casaponsa}, {Dickinson}, {Diego}, {Fern{\'a}ndez-Cobos}, {Grainge},
  {Guti{\'e}rrez}, {Herranz}, {Hoyland}, {Lasenby}, {L{\'o}pez-Caniego},
  {Mart{\'\i}nez-Gonz{\'a}lez}, {McCulloch}, {Melhuish}, {Piccirillo},
  {Perrott}, {Poidevin}, {Razavi-Ghods}, {Scott}, {Titterington}, {Tramonte},
  {Vielva}, \& {Vignaga}}]{QUIJOTE_I_2015}
{G{\'e}nova-Santos}, R., {Rubi{\~n}o-Mart{\'\i}n}, J.~A., {Rebolo}, R.,
  {et~al.} 2015, \mnras, 452, 4169

\bibitem[{{Kamionkowski} \& {Kovetz}(2016)}]{kamionkowski:2016}
{Kamionkowski}, M. \& {Kovetz}, E.~D. 2016, \araa, 54, 227

\bibitem[{{Keih{\"a}nen} {et~al.}(2005){Keih{\"a}nen}, {Kurki-Suonio}, \&
  {Poutanen}}]{keihanen2005}
{Keih{\"a}nen}, E., {Kurki-Suonio}, H., \& {Poutanen}, T. 2005, \mnras, 360,
  390

\bibitem[{{King} {et~al.}(2010){King}, {Copley}, {Davies}, {Davis},
  {Dickinson}, {Hafez}, {Holler}, {John}, {Jonas}, {Jones}, {Leahy},
  {Muchovej}, {Pearson}, {Readhead}, {Stevenson}, \& {Taylor}}]{king2010}
{King}, O.~G., {Copley}, C., {Davies}, R., {et~al.} 2010, in Society of
  Photo-Optical Instrumentation Engineers (SPIE) Conference Series, Vol. 7741,
  Society of Photo-Optical Instrumentation Engineers (SPIE) Conference Series,
  1

\bibitem[{{Minami} \& {Komatsu}(2020)}]{minami:2020}
{Minami}, Y. \& {Komatsu}, E. 2020, \prl, 125, 221301

\bibitem[{{Page} {et~al.}(2007){Page}, {Hinshaw}, {Komatsu}, {Nolta},
  {Spergel}, {Bennett}, {Barnes}, {Bean}, {Dor{\'e}}, {Dunkley}, {Halpern},
  {Hill}, {Jarosik}, {Kogut}, {Limon}, {Meyer}, {Odegard}, {Peiris}, {Tucker},
  {Verde}, {Weiland}, {Wollack}, \& {Wright}}]{page2007}
{Page}, L., {Hinshaw}, G., {Komatsu}, E., {et~al.} 2007, \apjs, 170, 335

\bibitem[{{Penzias} \& {Wilson}(1965)}]{penzias:1965}
{Penzias}, A.~A. \& {Wilson}, R.~W. 1965, \apj, 142, 419

\bibitem[{{\sorthelp{Planck Collaboration 2014A}}{Planck Collaboration
  I}(2014)}]{planck2013-p01}
{\sorthelp{Planck Collaboration 2014A}}{Planck Collaboration I}. 2014, \aap,
  571, A1

\bibitem[{{\sorthelp{Planck Collaboration 2014B}}{Planck Collaboration
  II}(2014)}]{planck2013-p02}
{\sorthelp{Planck Collaboration 2014B}}{Planck Collaboration II}. 2014, \aap,
  571, A2

\bibitem[{{\sorthelp{Planck Collaboration 2014I}}{Planck Collaboration
  IX}(2014)}]{planck2013-p03d}
{\sorthelp{Planck Collaboration 2014I}}{Planck Collaboration IX}. 2014, \aap,
  571, A9

\bibitem[{{\sorthelp{Planck Collaboration 2014L}}{Planck Collaboration
  XII}(2014)}]{planck2013-p06}
{\sorthelp{Planck Collaboration 2014L}}{Planck Collaboration XII}. 2014, \aap,
  571, A12

\bibitem[{{\sorthelp{Planck Collaboration 2015B}}{Planck Collaboration
  II}(2016)}]{planck2014-a03}
{\sorthelp{Planck Collaboration 2015B}}{Planck Collaboration II}. 2016, \aap,
  594, A2

\bibitem[{{\sorthelp{Planck Collaboration 2015C}}{Planck Collaboration
  III}(2016)}]{planck2014-a04}
{\sorthelp{Planck Collaboration 2015C}}{Planck Collaboration III}. 2016, \aap,
  594, A3

\bibitem[{{\sorthelp{Planck Collaboration 2015D}}{Planck Collaboration
  IV}(2016)}]{planck2014-a05}
{\sorthelp{Planck Collaboration 2015D}}{Planck Collaboration IV}. 2016, \aap,
  594, A4

\bibitem[{{\sorthelp{Planck Collaboration 2015J}}{Planck Collaboration
  X}(2016)}]{planck2014-a12}
{\sorthelp{Planck Collaboration 2015J}}{Planck Collaboration X}. 2016, \aap,
  594, A10

\bibitem[{{\sorthelp{Planck Collaboration 2018A}}{Planck Collaboration
  I}(2020)}]{planck2016-l01}
{\sorthelp{Planck Collaboration 2018A}}{Planck Collaboration I}. 2020, \aap,
  641, A1

\bibitem[{{\sorthelp{Planck Collaboration 2018B}}{Planck Collaboration
  II}(2020)}]{planck2016-l02}
{\sorthelp{Planck Collaboration 2018B}}{Planck Collaboration II}. 2020, \aap,
  641, A2

\bibitem[{{\sorthelp{Planck Collaboration 2018C}}{Planck Collaboration
  III}(2020)}]{planck2016-l03}
{\sorthelp{Planck Collaboration 2018C}}{Planck Collaboration III}. 2020, \aap,
  641, A3

\bibitem[{{\sorthelp{Planck Collaboration 2018D}}{Planck Collaboration
  IV}(2018)}]{planck2016-l04}
{\sorthelp{Planck Collaboration 2018D}}{Planck Collaboration IV}. 2018, \aap,
  641, A4

\bibitem[{{\sorthelp{Planck Collaboration IntZP}}{Planck Collaboration Int.
  XLI}(2016)}]{planck2015-XLI}
{\sorthelp{Planck Collaboration IntZP}}{Planck Collaboration Int. XLI}. 2016,
  \aap, 596, A102

\bibitem[{{\sorthelp{Planck Collaboration IntZZG}}{Planck Collaboration Int.
  LVII}(2020)}]{npipe}
{\sorthelp{Planck Collaboration IntZZG}}{Planck Collaboration Int. LVII}. 2020,
  \aap, 643, A42

\bibitem[{{Reinecke} {et~al.}(2006){Reinecke}, {Dolag}, {Hell}, {Bartelmann},
  \& {En{\ss}lin}}]{reinecke2006}
{Reinecke}, M., {Dolag}, K., {Hell}, R., {Bartelmann}, M., \& {En{\ss}lin},
  T.~A. 2006, \aap, 445, 373

\bibitem[{{Svalheim et al.}(2022)}]{bp14}
{Svalheim et al.} 2022, \aap, in preparation [\eprint[arXiv]{2011.08503}]

\bibitem[{{Tristram} {et~al.}(2021){Tristram}, {Banday}, {G{\'o}rski},
  {Keskitalo}, {Lawrence}, {Andersen}, {Barreiro}, {Borrill}, {Colombo},
  {Eriksen}, {Fernandez-Cobos}, {Kisner}, {Mart{\'\i}nez-Gonz{\'a}lez},
  {Partridge}, {Scott}, {Svalheim}, \& {Wehus}}]{tristram:2021}
{Tristram}, M., {Banday}, A.~J., {G{\'o}rski}, K.~M., {et~al.} 2021, arXiv
  e-prints, arXiv:2112.07961

\bibitem[{{Zonca} {et~al.}(2009){Zonca}, {Franceschet}, {Battaglia}, {Villa},
  {Mennella}, {D'Arcangelo}, {Silvestri}, {Bersanelli}, {Artal}, {Butler},
  {Cuttaia}, {Davis}, {Galeotta}, {Hughes}, {Jukkala}, {Kilpi{\"a}},
  {Laaninen}, {Mandolesi}, {Maris}, {Mendes}, {Sandri}, {Terenzi}, {Tuovinen},
  {Varis}, \& {Wilkinson}}]{zonca2009}
{Zonca}, A., {Franceschet}, C., {Battaglia}, P., {et~al.} 2009, Journal of
  Instrumentation, 4, 2010

\end{thebibliography}
\end{document}